\documentclass[twoside,twocolumn,9pt]{article}
\usepackage{extsizes}
\usepackage[super,sort&compress,comma]{natbib}
\usepackage[version=3]{mhchem}
\usepackage[left=1.5cm, right=1.5cm, top=1.785cm, bottom=2.0cm]{geometry}
\usepackage{balance}
\usepackage{times,mathptmx}
\usepackage{sectsty}
\usepackage{graphicx}
\usepackage{lastpage}
\usepackage[format=plain,justification=justified,singlelinecheck=false,font={stretch=1.125,small,sf},labelfont=bf,labelsep=space]{caption}
\usepackage{float}
\usepackage{fancyhdr}
\usepackage{fnpos}
\usepackage[english]{babel}
\usepackage{array}
\usepackage{droidsans}
\usepackage{charter}
\usepackage[T1]{fontenc}
\usepackage[usenames,dvipsnames]{xcolor}
\usepackage{setspace}
\usepackage[compact]{titlesec}
\usepackage{hyperref}
\usepackage{color}
\usepackage{amsmath}
\usepackage{amssymb}
\usepackage{bm}
\usepackage{multirow}
\usepackage{mathtools}

\definecolor{cream}{RGB}{222,217,201}

\begin{document}

\pagestyle{fancy}
\thispagestyle{plain}
\fancypagestyle{plain}{

\renewcommand{\headrulewidth}{0pt}
}
\makeFNbottom
\makeatletter
\renewcommand\LARGE{\@setfontsize\LARGE{15pt}{17}}
\renewcommand\Large{\@setfontsize\Large{12pt}{14}}
\renewcommand\large{\@setfontsize\large{10pt}{12}}
\renewcommand\footnotesize{\@setfontsize\footnotesize{7pt}{10}}
\makeatother

\renewcommand{\thefootnote}{\fnsymbol{footnote}}
\renewcommand\footnoterule{\vspace*{1pt}%
\color{cream}\hrule width 3.5in height 0.4pt \color{black}\vspace*{5pt}}
\setcounter{secnumdepth}{5}

\makeatletter
\renewcommand\@biblabel[1]{#1}
\renewcommand\@makefntext[1]%
{\noindent\makebox[0pt][r]{\@thefnmark\,}#1}
\makeatother
\renewcommand{\figurename}{\small{Fig.}~}
\sectionfont{\sffamily\Large}
\subsectionfont{\normalsize}
\subsubsectionfont{\bf}
\setstretch{1.125} 
\setlength{\skip\footins}{0.8cm}
\setlength{\footnotesep}{0.25cm}
\setlength{\jot}{10pt}
\titlespacing*{\section}{0pt}{4pt}{4pt}
\titlespacing*{\subsection}{0pt}{15pt}{1pt}

\fancyfoot{}
\fancyfoot[RO]{\footnotesize{\sffamily{\thepage}}}
\fancyfoot[LE]{\footnotesize{\sffamily{\thepage}}}
\fancyhead{}
\renewcommand{\headrulewidth}{0pt}
\renewcommand{\footrulewidth}{0pt}
\setlength{\arrayrulewidth}{1pt}
\setlength{\columnsep}{6.5mm}
\setlength\bibsep{1pt}

\makeatletter
\newlength{\figrulesep}
\setlength{\figrulesep}{0.5\textfloatsep}

\newcommand{\topfigrule}{\vspace*{-1pt}%
\noindent{\color{cream}\rule[-\figrulesep]{\columnwidth}{1.5pt}} }

\newcommand{\botfigrule}{\vspace*{-2pt}%
\noindent{\color{cream}\rule[\figrulesep]{\columnwidth}{1.5pt}} }

\newcommand{\dblfigrule}{\vspace*{-1pt}%
\noindent{\color{cream}\rule[-\figrulesep]{\textwidth}{1.5pt}} }

\makeatother

\title{\textbf{Mesoscopic Modelling and Simulation of Soft Matter}}
\date{}
\author{\large Ulf D. Schiller~\textit{$^{a}$}, Timm Kr\"uger~\textit{$^{b}$} and Oliver Henrich~$^{\ast}$\textit{$^{c}$}}
\maketitle

\abstract{\normalsize \it The deformability of soft condensed matter often requires 
modelling of hydrodynamical aspects to gain quantitative understanding. This, however, requires specialised 
methods that can resolve the multiscale nature of soft matter systems. We review a number of the most popular simulation methods that 
have emerged, such as Langevin dynamics, dissipative particle dynamics, multi-particle collision dynamics, sometimes also 
referred to as stochastic rotation dynamics, and the lattice-Boltzmann method. We conclude this review with a short glance at 
current compute architectures for high-performance computing and community codes for soft matter simulation.}


\renewcommand*\rmdefault{bch}\normalfont\upshape
\rmfamily
\section*{}
\vspace{-1cm}


\footnotetext{\textit{$^{a}$~Department of Materials Science and Engineering, Clemson University, 161 Sirrine Hall, Clemson, SC 29634, USA}}
\footnotetext{\textit{$^{b}$~School of Engineering, University of Edinburgh, Edinburgh EH9 3JL, Scotland, UK}}
\footnotetext{\textit{$^{c}$~SUPA, Department of Physics, University of Strathclyde, Glasgow G4 0NG, Scotland, UK, E-mail: oliver.henrich@strath.ac.uk}}




\newcommand{\e}[1]{\times10^{#1}}
\newcommand{\beq}{\begin{equation}}
\newcommand{\eeq}{\end{equation}}
\newcommand{\beqa}{\begin{eqnarray}}
\newcommand{\eeqa}{\end{eqnarray}}
\newcommand{\com}[1]{\textcolor{red}{#1}}
\newcommand{\rev}[1]{{#1}}
\newcommand{\cur}[1]{{\textit{#1}}}
\newcommand{\ben}{\begin{enumerate}}
\newcommand{\een}{\end{enumerate}}
\newcommand{\bit}{\begin{itemize}}
\newcommand{\eit}{\end{itemize}}

\pagestyle{fancy}


\renewcommand{\vec}[1]{{\ensuremath{\mathchoice
                     {\mbox{\boldmath$\displaystyle\mathbf{#1}$}}
                     {\mbox{\boldmath$\textstyle\mathbf{#1}$}}
                     {\mbox{\boldmath$\scriptstyle\mathbf{#1}$}}
                     {\mbox{\boldmath$\scriptscriptstyle\mathbf{#1}$}}}}}%

\newcommand{\tens}[1]{{\ensuremath{\mathchoice
                     {\mbox{$\displaystyle\mathsf{#1}$}}
                     {\mbox{$\textstyle\mathsf{#1}$}}
                     {\mbox{$\scriptstyle\mathsf{#1}$}}
                     {\mbox{$\scriptscriptstyle\mathsf{#1}$}}}}}%

\newcommand{\jvec}{\ensuremath{\vec{j}}}
\newcommand{\rvec}{\ensuremath{\vec{r}}}
\newcommand{\uvec}{\ensuremath{\vec{u}}}
\newcommand{\vvec}{\ensuremath{\vec{v}}}
\newcommand{\xvec}{\ensuremath{\vec{x}}}

\vspace*{0.5cm}

\section{Introduction}

Soft condensed matter \cite{Doi:2013, Terentjev:2015} has an ubiquitous presence in our world and we 
encounter it in our everyday lives. Many complex fluids like polymer solutions and colloidal suspensions, 
liquid crystals, foams, gels, granular materials and biological materials belong to this category. 
A feature that distinguishes soft matter from more conventional condensed matter 
is that it shows a remarkable propensity to self-organise and form more complex, multiscale structures
which exist on intermediate mesoscopic time and length scales much larger than the atomistic
length scale, but also much smaller than the macroscopic lab scale. 
This characteristic nature of soft matter entails a typical separation of time and length scales and becomes
important for solvent-mediated interactions e.g. between suspended nanoparticles or in systems with 
internal degrees of freedom like fluctuating membranes, polymer chains or vesicles. Thus, it can be quite 
challenging to find a consistent physical description that covers all relevant aspects of the problem.

Nonlinear coupling mechanisms between different components of the physical model, such as the order-flow coupling
in liquid crystals, occur frequently in soft matter. Sometimes they prevent accurate analytical solutions 
and make simulations more favourable, or even outright indispensable.
For a quantitative understanding of the response and dynamical behaviour it is 
crucial to find a suitable coarse-grained descriptions that manages to express a large number
of degrees of freedom through a much smaller number of effective degrees of freedom whilst retaining
the correct overall physical behaviour, therefore allowing to bridge time and length scales.
An example is flowing soft matter. Quite generally speaking, flow is of particular importance due
to the deformability of soft matter. Usually it can be assumed the flow is incompressible and characterised
by low Reynolds numbers. Hydrodynamic interactions, however, lead to long-ranged, collective interactions 
that are notoriously difficult to treat with analytical models. 

These examples highlight only a few complications on the way to model soft matter. They have led to the 
formulation of specialised simulation methods which are the focus of this tutorial review.  
Due to space limitations we can cover here unfortunately only the most common and versatile methods.
In the following article, we will present a short overview of these relatively recent developments and 
glance also at some typical applications of these simulation methods and make suggestions for further reading.

\subsection{Mesoscopic Modelling: Particle-based vs. Lattice Models}

The mesoscopic methods discussed in this review are essentially alternative ways to model the dynamic correlations between solute particles that are mediated by momentum transport in a solvent medium. The momentum transport in the solvent is in principle described by the Navier-Stokes equation, however, the dynamics can also be affected by thermal fluctuations and specific molecular-level interactions. Mesoscopic methods are based on ``coarse-graining'' the microscopic details, i.e., including only the essential details of the interactions thus greatly reducing the degrees of freedom of the system. For instance, the hydrodynamic interactions between solute particles are amenable to a Langevin description
\begin{equation}\label{eq:brownian-dynamics}
\dot{\vec{r}}_i = \sum_j \frac{\mathsf{D}_{ij}}{k_BT} \vec{F}_j + \Delta\vec{r}_i ,
\end{equation}
%
%
where $\vec{F}_j$ is the effective conservative force acting on particle $j$, $\tens{D}_{ij}$ is the diffusion tensor, and $\Delta\vec{r}_i$ are stochastic displacements \rev{that represent thermal fluctuations and satisfy the fluctuation dissipation relation
\begin{align}
\left\langle \Delta\vec{r}_i \right\rangle &= 0 , \\
\left\langle \Delta\vec{r}_i(t) \Delta\vec{r}_j(t') \right\rangle &= 2 \mathsf{D}_{ij} \delta(t-t') .
\end{align}
}
In soft matter, diffusion of the solutes is much slower than diffusion of momentum leading to Schmidt numbers $Sc=\nu/D$ on the order of $10^6$ for micron size particles. Any mesoscopic method should therefore guarantee $Sc \gg 1$. The diffusion tensor depends on the position of all particles in the system, and a simplified, pair-wise additive form is given by the Rotne-Prager tensor\cite{Rotne:1969}
%
%
\rev{
\begin{align}\label{eq:rotne-prager}
\tens{D}_{ij} &= D_0\; \Bigg\{\left(1-\frac{9\,r_{ij}}{32\,a}\right) \mathbf{I} + \frac{3}{32} \frac{\mathbf{r}_{ij} \otimes \mathbf{r}_{ij}}{a\,r_{ij}}\Bigg\}, &\quad r_{ij} < 2a\nonumber\\
\tens{D}_{ij} &=\nonumber
D_0\; \frac{3\,a}{4\,r_{ij}}\;\Bigg\{\mathbf{I} + \frac{\mathbf{r}_{ij} \otimes \mathbf{r}_{ij}}{r^2_{ij}} \\ &\qquad\qquad\qquad+ \frac{2\,a^2}{3\,r^2_{ij}} \left(\mathbf{I} - 3 \frac{\mathbf{r}_{ij} \otimes \mathbf{r}_{ij}}{r^2_{ij}}\right)\Bigg\}, &\quad r_{ij}\ge 2a\nonumber\\
\tens{D}_{ii} &= D_0\; \mathbf{I}
\end{align}
}
where $a$ is the radius of the suspended particles and $\vec{r}_{ij}=\vec{r}_i-\vec{r}_j$.
The equation of motion Eq. (\ref{eq:brownian-dynamics}) can in principle be integrated numerically, and the most straightforward approach is known as Brownian dynamics \cite{Ricci:2003}.
%
%
\rev{The time evolution of the system can accordingly be written
\begin{equation}
\vec{r}_i(t+h) = \vec{r}_i(t) + \frac{\mathsf{D}_{ij}}{k_BT} h \vec{F}_j + \sqrt{2 h} \, \mathsf{B}_{ij} \cdot \vec{W}_j ,
\end{equation}
where $\vec{W}_j$ are random vectors representing a discretised Wiener process such that $\left\langle \vec{W}_i \right\rangle = 0$ and $\left\langle \vec{W}_i \otimes \vec{W}_j \right\rangle = \mathbf{I} \delta_{ij}$. The tensor $\mathsf{B}_{ij}$ is related to the diffusion tensor through $\mathsf{D}_{ij} = \mathsf{B}_{ik}\cdot\mathsf{B}^T_{jk}$ and can be represented by a Cholesky decomposition into an upper triangular matrix $\mathsf{C}_{ij}$ such that $\mathsf{D}_{ij}=\mathsf{C}_{ik}\cdot\mathsf{C}^T_{jk}$. The use of the Cholesky factorisation for BD was proposed by Ermak and McCammon \cite{Ermak:1978}, however, the algorithm scales as $O(N^3)$ with the particle number $N$ and is infeasible already for a few hundred particles. Fixman \cite{Fixman:1986b} introduced a more efficient procedure using a Chebyshev polynomial approximation
\begin{equation}
  \mathsf{B}_{ij} = \sum_{l=0}^L a_l C_l\left( \mathsf{E} \right) - \frac{a_0}{2} C_0\left(\mathsf{E}\right) ,
\end{equation}
where $\mathsf{E}=(\mathsf{D}_{ij}-h_+)/h_-$, $h_+=(\lambda_\text{max}+\lambda_\text{min})/2$ and $h_-=(\lambda_\text{max}-\lambda_\text{min})/2$, and $\lambda_\text{max}$ and $\lambda_\text{min}$ are the largest and smallest eigenvalue of the diffusion tensor, respectively. The Chebyshev polynomials $C_l(\mathsf{E})$ are given by the recursion relation
\begin{align}
  C_0(\mathsf{E}) &= \mathbf{I} \\
  C_1(\mathsf{E}) &= \mathsf{E} \\
  C_{l+1}(\mathsf{E}) &= 2 \mathsf{E} \cdot C_l(\mathsf{E}) - C_{l-1}(\mathsf{E}) .
\end{align}
The coefficients $a_l$ are obtained from the Chebyshev series expansion of the square root function $f(e)=\sqrt{h_+ + h_- e}$
\begin{equation}
a_l = \frac{2}{L+1} \sum_{k=0}^{L} \cos\left( \frac{\pi l (k+\frac{1}{2})}{L+1} \right) f \left[ \cos \left( \frac{\pi(k+\frac{1}{2})}{L+1} \right) \right]^\frac{1}{2} .
\end{equation}
Instead of solving for $\mathsf{B}_{ij}$ directly, Fixman's algorithm uses the polynomial expansion is used to calculate the stochastic displacement
\begin{equation}
  \mathsf{B}_{ij} \vec{W}_j = \sum_{l=0}^{L} a_l \vec{x}_l -\frac{a_0}{2} \vec{x}_0 ,
\end{equation}
where $\vec{x}_l=C_l((\mathsf{D}_{ij}-h_+)/h_-)\cdot\vec{W}_j$.
%
%
The Chebyshev polynomial approximation of $\mathsf{B}_{ij}\cdot\vec{W}_j$ for a given accuracy scales roughly as $O(N^{2.25})$ which can be reduced to $O(N^2)$ by truncated Chebyshev expansions. For further details and a comparison of different implementations we refer the reader to Ref. \citenum{Schmidt:2011}.
The Chebyshev expansion may be sufficient to treat on the order of $10^3$ particles, however, in practice the runtime behaviour depends strongly on the desired accuracy and the details of the underlying physics. For polymer chains, Schmidt et al.\cite{Schmidt:2011} have found that the performance of Chebyshev-based procedures is significantly affected by overlap of the chains. An advantage of BD methods is that they do not rely on an underlying simulation box and thus do not exhibit finite box size effects. Hence, for single chains, BD methods may be superior to explicit solvent methods due to the $L^3$ scaling of the fluid degrees of freedom. However, for semi-dilute polymer solutions, Pham et al. \cite{Pham:2009} have estimated that the scaling ratio between BD and explicit solvent (e.g. LB) methods will tip in favour of the latter. Larger BD systems thus require even faster algorithms such as Ewald-like methods using fast Fourier transformations \cite{Banchio:2003}. These methods scale as $O(N^{1+x}\log N)$ where $x$ is typically substantially smaller than unity. However, these methods require the study of confined systems and cannot easily be generalised to arbitrary boundary conditions. Moreover, the underlying Langevin description of BD neglects the retardation effect associated with the finite speed of momentum propagation in the solvent.
}

In this review, we therefore focus on mesoscopic methods that maintain a coarse-grained description of the solvent with explicit momentum transport. One can distinguish two broad classes of mesoscopic methods, namely particle-based and lattice models. Particle based-methods, such as dissipative particle dynamics (DPD) and multi-particle collision dynamics (MPC), represent the solvent by a system of interacting particles. The ``coarse-graining'' of the molecular details is achieved by implementing the interactions through collective collisions that satisfy the local conservation laws. Particle-based methods maintain a continuous phase space and thermal fluctuations are inherently present. DPD is essentially a momentum-conserving version of the Langevin thermostat and its algorithm is closely related to molecular dynamics based on Newton's equations of motion. In contrast, MPC is not developed as a time-discrete scheme for integrating a continuous equation of motion, but is based on discrete streaming and collision steps similar to Bird's direct simulation Monte Carlo (DSMC) method\cite{Bird:1994}. Both DPD and MPC can be shown to satisfy an $H$-theorem\cite{Malevanets:1999,Ihle:2003a,Marsh:1998}. In addition, MPC can easily be switched from a micro-canonical ensemble to a canonical ensemble by augmenting the collision rule with \rev{a thermostat\cite{Bolintineanu:2012,Huang:2015}.}

Lattice models, such as finite-element models or the lattice Boltzmann method (LBM) represent the solvent by hydrodynamic fields on a discrete lattice. Thermal fluctuations can be reintroduced by means of stochastic collisions, if needed. In lattice models, the solvent viscosity (and other transport coefficients) are directly linked to simulation parameters and thus can be set without the need for calibration. Moreover, the LBM can be rigorously derived from kinetic theory which provides a systematic route extensions of its applicability to, e.g., multiphase fluids or Knudsen flows \cite{Yeomans:2006,Toschi:2005}.

Both particle-based and lattice mesoscopic methods define a discrete dynamics that can be shown to reproduce Navier-Stokes hydrodynamics asymptotically. In this sense, the parameters of the coarse-grained model represent constitutive relations on the macroscopic level that can be tuned to reproduce the transport properties of the real physical system, cf.~section \ref{sec:parameter-choice}. Moreover, the methods allow implementation of boundary conditions and provide systematic means of coupling solute particles to the solvent medium. For these reasons, mesoscopic methods are perfectly suited to study complex phenomena in soft matter systems, both in and out of equilibrium.

\subsection{Parameter Choice in Mesoscopic Simulations}
\label{sec:parameter-choice}

A crucial step in any mesoscopic simulation is the mapping of simulation parameters to physical quantities. Whereas many authors resort to what is commonly called ``simulation units'', there is often confusion as to how these are set and sometimes the details provided are insufficient to reproduce the results. 
Mesoscopic methods provide some flexibility in choosing a parameter mapping in such a way that the physical properties of interest are correctly captured by means of \emph{dimensionless numbers}.
The hydrodynamic transport properties of fluids are characterised by dimensionless numbers including the Schmidt number $Sc$, the Mach number $Ma$, the Reynolds number $Re$, the Knudsen number $Kn$, and the P{\'e}clet number $Pe$:
\begin{align}
Sc &= \frac{\nu}{D_f}, \\
Ma &= \frac{u}{c_s}, \\
Re &= \frac{uL}{\nu} = \frac{c_s L}{\nu} Ma, \\
Kn &= \frac{\lambda}{L} \propto \frac{Ma}{Re}, \\
Pe &= \frac{uL}{D_s} = \frac{\nu}{D_s} Re = \frac{c_sL}{D_s} Ma .
\end{align}

Here, $\lambda$ is the mean free path, $D_f$ is the self-diffusion coefficient of the solvent, and $D_s$ is the diffusion coefficient of the solute particles. The value of $D_s$ typically depends on the details of the coupling to the fluid for the respective method. In particle-based methods, a large Schmidt number can only be obtained if the mean free path is small, which usually requires a small time step $h$. This is a common characteristic of coarse-grained particle-based simulations that use a reduced number of particles which leads to a larger mean free path. In practice it is typically sufficient to ensure that the momentum transport is fast enough to generate liquid-like behaviour, and simulations can still be carried out with Schmidt numbers on the order of $\mathcal{O}(1)$ to $\mathcal{O}(10)$. Similarly, it is feasible to simulate at a higher Mach number than in the real system, as long as density fluctuations remain sufficiently small (incompressible limit).

Arguably the most important dimensionless number characterising hydrodynamic flow is the Reynolds number that quantifies the ratio of inertial and viscous momentum transport. In simulations of soft matter and microflows, the Reynolds number is typically small such that nonlinear inertial effects do not play an important role. The flow velocity is usually a result of external fields, and hence the Reynolds number can be used to tune the parameters that determine the magnitude of the driving forces of the simulation in relation to the viscosity of the fluid.

In terms of standard dimensionless groups, the ratio of the Mach number and the Reynolds number is proportional to the Knudsen number that quantifies the importance of rarefaction effects that can occur in microflows and lead to deviations from continuum Navier-Stokes behaviour. Knudsen numbers beyond $Kn\gtrsim 0.1$ indicate that the flow is in the transition regime where non-continuum effects can become significant. 
The P{\'e}clet number quantifies the relative importance of convective and diffusive transport of solutes which is related to the importance of thermal fluctuations, i.e., for small P{\'e}clet numbers Brownian motion dominates hydrodynamic advection. 
In microflows, the ratio $Pe/Re$ is typically very large and it may not be feasible to use a sufficiently large number of particles to reproduce the P{\'e}clet number. In such systems one can seek a compromise and simulate at a higher Reynolds number and/or lower P{\'e}clet number, as long as one stays in the correct hydrodynamic regime which should be carefully validated.

In order to determine the simulation parameters, it is also instructive to consider the time scales of interest\cite{Padding:2006}. The main hydrodynamic time scales are the acoustic time scale $\tau_{c_s} = L/c_s$, the viscous (kinematic) time scale $\tau_\nu = L^2/\nu$, the diffusive time scale $\tau_D = L^2/D_s$, and the Stokes time scale $\tau_\text{S} = L/u$. These time scales can be related to each other using dimensionless numbers
\begin{equation}
\tau_\text{S} = \frac{\tau_{D_s}}{Pe} = \frac{\tau_\nu}{Re} = \frac{\tau_{c_s}}{Ma} .
\end{equation}
This equation makes it evident how the separation between the time scales depends on the dimensionless numbers. It may thus not always be possible (and desirable) to resolve all time scales in the same simulation. In particular, the separation between the viscous and the acoustic time scale is typically large, which can be exploited to simulate the system with a speed of sound that is smaller than the real physical one.

\section{Dissipative Particle Dynamics}

Dissipative particle dynamics (DPD) is a stochastic simulation method that was specifically designed for soft matter
and complex fluids. It was first formulated by Hoogerbrugge and Koelman 
\cite{Hoogerbrugge:1992, Koelman:1993} and later refined by Espa\~nol \cite{Espanol:1995b} and 
Groot and Warren \cite{Groot:1997}.

\subsection{Basic Algorithm}

The underlying idea is akin to coarse-grained molecular dynamics with atoms agglomerated into larger
entities or ``beads'' that interact via soft forces. 
These beads are subject to conservative forces as well as pairwise drag
or friction forces and random forces. The force balance for bead $i$ is slightly different than in Eq.~\ref{eq:brownian-dynamics} as it 
runs over particle pairs $(i,j)$:
\beq\label{ftotal}
m_i \ddot{{\mathbf{r}}}_i=\mathbf{F}_i^\text{C}+\sum_{j\neq i}\mathbf{F}_{i j}^\text{D} + \sum_{j\neq i}\mathbf{F}_{i j}^\text{R}.
\eeq
The functional forms of the forces are also slightly different. For instance, the conservative force 
\beq\label{fconservative}
\mathbf{F}_i^\text{C} (\{\mathbf{r}\})= -\frac{\partial V (\{\mathbf{r}\})}{\partial\mathbf{r}_i}
\eeq
always contains a soft-core repulsion in addition to other interactions or external forces.
 
The drag force depends on the relative separation $\mathbf{r}_{ij} = \mathbf{r}_i - \mathbf{r}_j$ and velocity $\mathbf{v}_{ij} = \mathbf{v}_i - \mathbf{v}_j$ of the beads,
\beq\label{fdrag}
\mathbf{F}_{ij}^\text{D} (\mathbf{r}_{ij},\mathbf{v}_{ij})=-\gamma\,\omega(|\mathbf{r}_{ij}|)\frac{(\mathbf{r}_{ij}\cdot\mathbf{v}_{ij})\mathbf{r}_{ij}}{|\mathbf{r}_{ij}|},
\eeq
whereas the random forces are given by
\beq\label{frandom}
\mathbf{F}_{ij}^\text{R} (\mathbf{r}_{ij})=\sqrt{2 k_B T \gamma\, \omega(|\mathbf{r}_{ij}|)} \frac{\mathbf{r}_{ij}}{|\mathbf{r}_{ij}|}\, \xi_{ij}.
\eeq
The quantity $\xi_{i j}$ is a Gaussian random number with zero mean and unit variance which is symmetric with respect to the particle 
indices, i.e., $\xi_{i j} = \xi_{j i}$. This is a requirement for total momentum conservation and in contrast to Brownian and Langevin dynamics where
the stochastic noise on each particle is independent of all other particles. $\gamma$ is a friction coefficient and
$\omega(|\mathbf{r}_{ij}|)$ a decaying weighing function with a model-specific cutoff distance \cite{Moeendarbary:2009, Moeendarbary:2010}.
In order to fulfil a fluctuation-dissipation theorem in DPD, distance-dependent friction forces require distance-dependent random forces for each pair of particles.
This requirement has been demonstrated at Gibbsian equilibrium by Espa\~nol and Warren \cite{Espanol:1995a}.

Various schemes have been proposed to perform the time integration of Eq.~\ref{ftotal} \cite{Nikunen:2003}.
The simplest integrator that can be considered a bare minimum requirement is based on a modified velocity-Verlet algorithm:

\beqa
\mathbf{v}_i(t+h/2)&=&\mathbf{v}_i(t) + \frac{\sqrt{h}}{2 m_i} \sum_{j\neq i} \mathbf{F}_{ij}^\text{R}(\mathbf{r}_{ij}(t))\nonumber\\[-15pt]
& + & \frac{h}{2 m_i} \left(\mathbf{F}_i^\text{C}(\{\mathbf{r}(t)\}) + \sum_{j\neq i} \mathbf{F}_{ij}^\text{D} (\mathbf{r}_{ij}(t), \mathbf{v}_{ij}(t))\right), \label{vv1}\\
\mathbf{r}_i(t+h)&=& \mathbf{r}_i(t) + h\,\mathbf{v}_i(t+h/2), \label{vv2}\\ 
\mathbf{v}^{\star}_i(t+h)&=& \mathbf{v}_i(t+h/2) + \frac{\sqrt{h}}{2 m_i} \sum_{j\neq i} \mathbf{F}_{ij}^\text{R}(\mathbf{r}_{ij}(t+h))\nonumber\\[-12pt]
& + & \frac{h}{2 m_i}  \mathbf{F}_i^\text{C}(\{\mathbf{r}(t+h)\}), \label{vv3}\\
\mathbf{v}_i(t+h)&=&\mathbf{v}^{\star}_i(t+h) \nonumber\\[-5pt]
& + & \frac{h}{2 m_i} \sum_{j\neq i} \mathbf{F}_{ij}^\text{D} (\mathbf{r}_{ij}(t+h), \mathbf{v}_{ij}(t+h/2)). \label{vv4}
\eeqa

Note the square root of the time step $h$ which appears in Eq.~\ref{vv1} and Eq.~\ref{vv3}
when the stochastic Wiener process is discretised.
In DPD the velocities $\mathbf{v}_i$ at the end of the time step 
depend on the drag forces $\mathbf{F}_{ij}^\text{D}$ which 
in turn depend on the relative velocities. Hence, rather than just taking the drag forces based on the 
intermediate velocities at the half time step, a number of flavours of the DPD algorithm 
solve Eq.~\ref{vv4} in a self-consistent manner. In its simplest form
the drag forces are recalculated once using the velocities $\mathbf{v}_i(t+h)$ as obtained in Eq.~\ref{vv4}
and are then used in a final update of the velocities at $t+h$.
This improves the performance significantly.

The local and pairwise interactions in DPD fulfil Newton's third law, conserve momentum and angular momentum, 
guarantee Galilean invariance and yield hydrodynamic conservation laws on larger length scales owing to the 
particle-based nature of the algorithm.
A version of DPD with energy conservation can also be formulated  \cite{Espanol:1997}.
Together with a modified predictor-corrector algorithm for the integration \cite{Groot:1997} or a self-consistent
velocity-Verlet algorithm \cite{Pagonabarraga:2001}, DPD permits using larger time steps than atomistic MD 
modelling.

\subsection{Model Extensions}

A focus of interest has been around the thermodynamic consistency of DPD and free energy functionals \cite{Pagonabarraga:2001}.
If different species are modelled, the correct compressibility and solubility of the components, 
specified by the repulsion parameters between the different species, has to be provided \cite{Groot:1997}.
Otherwise there is considerable freedom in modelling the interactions.
DPD can also be derived from a coarse-grained version of molecular dynamics 
\cite{Flekkoy:1999, Flekkoy:2000}. It is even possible to establish a link
between DPD and a formulation of smoothed particle hydrodynamics (SPH) \cite{Espanol:2003}.
The resulting ``smoothed'' formulation of DPD is thermodynamically consistent 
and allows arbitrary equations of state.

Because pairwise interactions have to be calculated, DPD is computationally relatively expensive and  
``slower'' than other methods, such as multi-particle collision dynamics or the lattice-Boltzmann method.
Another disadvantage of DPD is that momentum transport is tightly coupled to particle transport and the Schmidt number 
is typically very low. However, schemes for arbitrarily large Schmidt numbers have been presented \cite{Lowe:2004}.

Further details on DPD can be found in the reviews by Nielsen \cite{Nielsen:2004}, 
Moeendarbary \cite{Moeendarbary:2009, Moeendarbary:2010} and Liu \cite{liu_dissipative_2015}.

\subsection{Applications}

The DPD method has been used to model different dynamic regimes in polymer melts \cite{Nikunen2007}.
Here, the challenge lies in finding a coherent description for the crossover from Rouse dynamics of freely moving chains 
undergoing hydrodynamic interaction at short chain lengths to reptational dynamics at long chain lengths, where the 
individual chains are entangled and feel the topological constraints formed by other chains in their vicinity. 
Because of the softness of the beads, standard DPD models are unable to prevent unphysical bond crossings and 
disentanglement of the polymer chains. It is therefore necessary to apply the right degree of
coarse graining by choosing the correct bead size and adapt the bond stretching interactions 
between the beads in such a way that disentanglement does not occur.

Another application area are electrohydrodynamic effects such as electro-osmosis and electrophoresis \cite{Smiatek:2011}.
The method was recently used to model slip boundary conditions close to hydrophobic surfaces. Interestingly, 
it was shown that both confinement and mobility of the surface charges have a dramatic effect on the hydrodynamic properties 
of the electric double layer and the electro-osmotic flow\cite{Maduar:2015}.

\begin{figure}[h]
 \centering
 \includegraphics[width=.9\columnwidth]{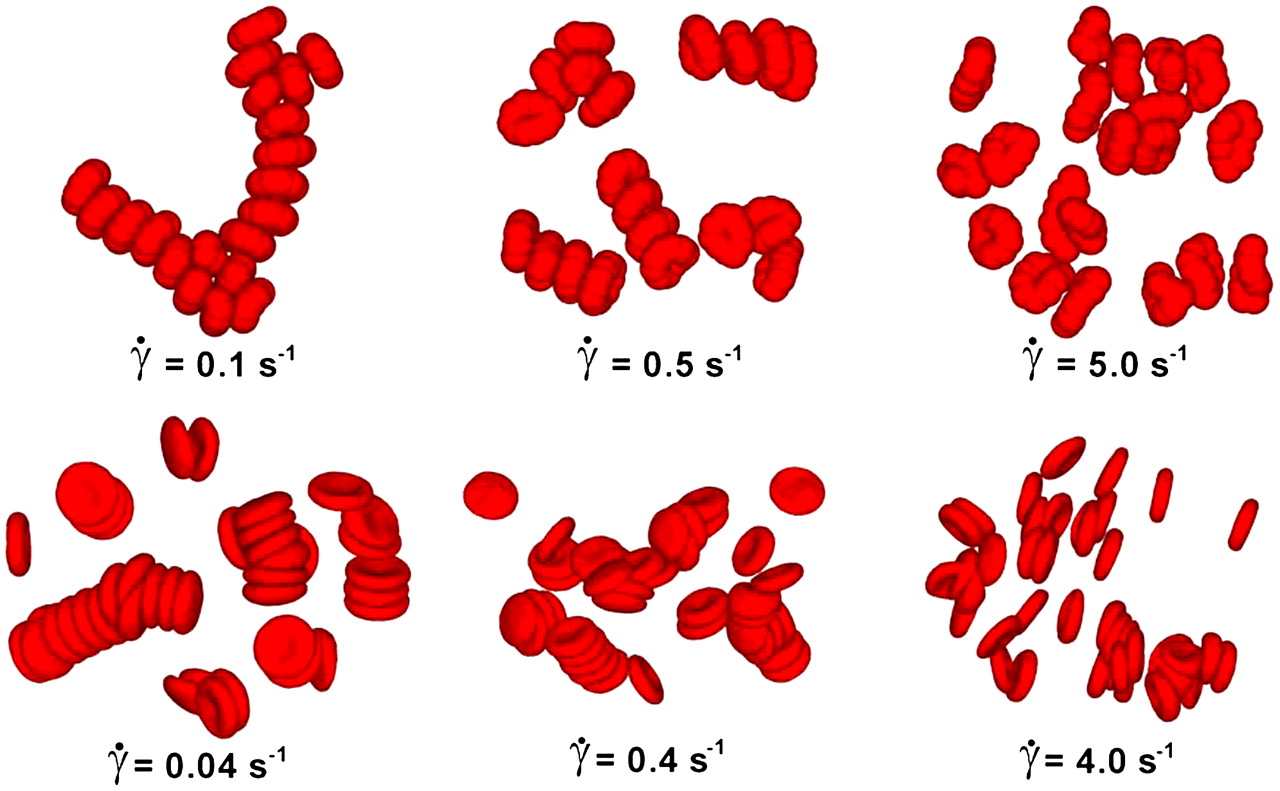}
 \caption{Visualisation of aggregation of red blood cells. Simulated reversible rouleaux are formed by a low-dimensional model (upper) and multiscale model (lower). The left column corresponds to low shear rates, centre column to moderate share rates, and right column to high shear rates \cite{fedosov_predicting_2011}. Copyright (2011) National Academy of Sciences.}
 \label{fig:dpd-blood}
\end{figure}

DPD has been employed to simulate the dynamics and rheology of red blood cells (RBCs) suspended in blood plasma, cf.~Fig.~\ref{fig:dpd-blood} \cite{fedosov_multiscale_2010, fedosov_predicting_2011}.
These cells are essentially biconcave viscoelastic shells (membranes) made of a lipid bilayer and a cytoskeleton, filled with a Newtonian haemoglobin solution.
Numerically, all fluid and solid components are discretised as DPD particles: the external blood plasma with a viscosity of about $\eta_\text{ex} \approx 1.2~\text{mPa s}$, the internal haemoglobin solution ($\eta_\text{in} \approx 6~\text{mPa s}$), and the RBC membrane as a collection of DPD particles subject to additional viscoelastic forces.
The membrane particles at locations $\{\mathbf{r}_i\}$ form a triangular mesh and have the potential
\beq
 V(\{\mathbf{r}_i\}) = V_\text{plane} + V_\text{bend} + V_\text{surf} + V_\text{vol}
\eeq
where $V_\text{plane}$ is the in-plane shear contribution due to the RBC cytoskeleton, $V_\text{bend}$ captures the bending resistance of the lipid bilayer, and $V_\text{surf}$ and $V_\text{vol}$ provide near conservation of the RBC surface area and volume which result from the incompressibility and impermeability of the RBC membrane.
Additionally, in-plane viscous forces are taken into account that capture the viscosity of the lipid bilayer.
Both external and internal liquid DPD particles are bounced back at either side of the membrane surface to ensure impermeability of the RBC membrane.
The dissipative forces between membrane and liquid particles are specifically chosen to enforce the no-slip boundary condition.


\section{Multi-Particle Collision Dynamics}

Multi-particle collision dynamics (MPC), originally introduced by Malevanets and Kapral\cite{Malevanets:1999} as stochastic rotation dynamics (SRD) is a particle-based mesoscopic method that has become popular in the soft matter domain thanks to the flexibility in handling spatiotemporally varying forces. MPC is well suited to study complex phenomena in soft matter both in and out of equilibrium. In the following, we summarise the essential elements of MPC. A comprehensive review of MPC was published by Gompper et al.\cite{Gompper:2009}

\subsection{Algorithm}

\begin{figure}[h]
  \centering
  \includegraphics[width=.9\columnwidth]{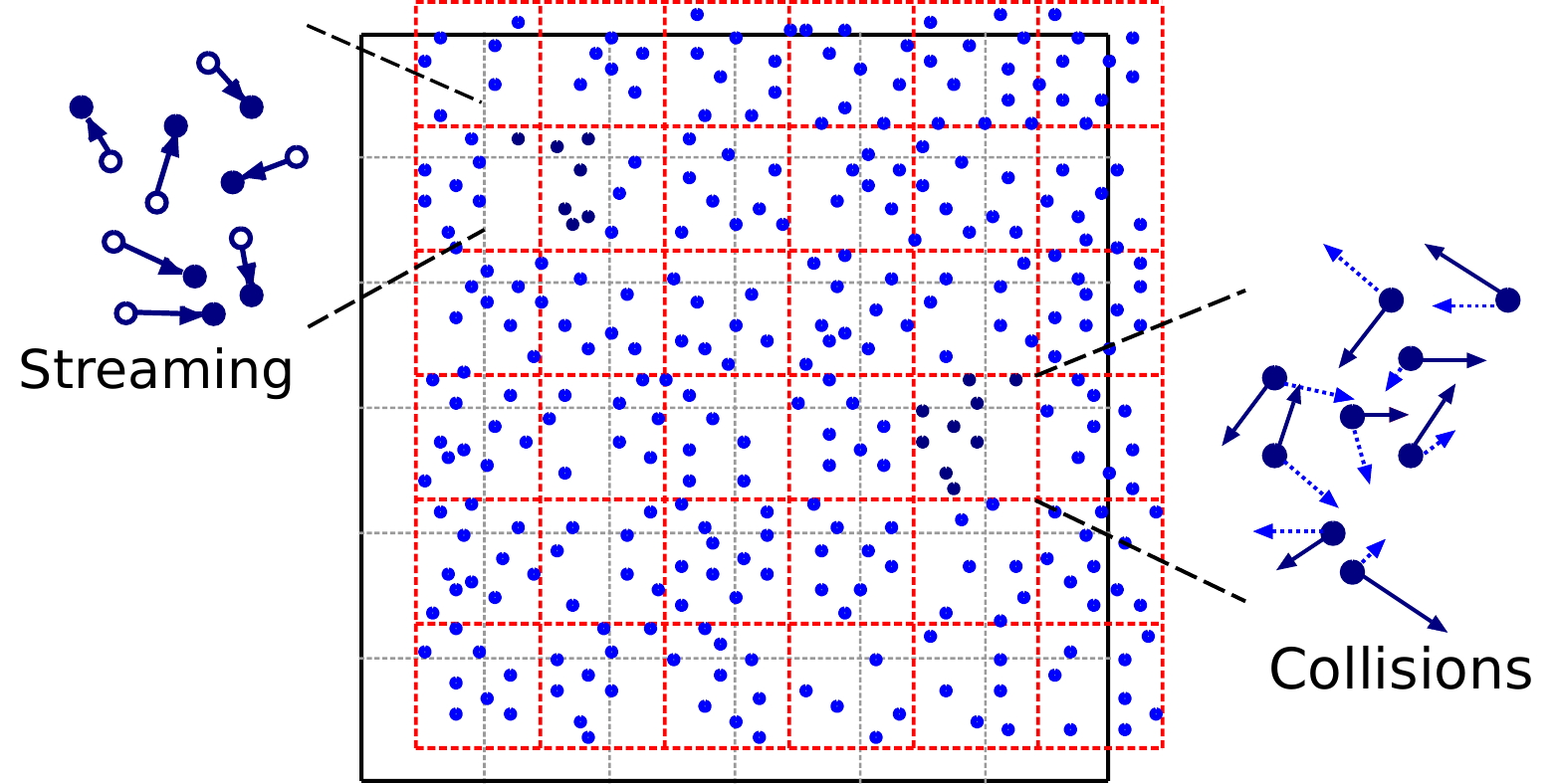}
  \caption{The basic algorithm of multi-particle collision dynamics consists of successive streaming and collision steps. 
The cell grid is shifted randomly every time step to restore Galilean invariance\cite{Ihle:2001,Ihle:2003a}.}
  \label{fig:mpc-algorithm}
\end{figure}

In MPC, the fluid consists of idealised point-like particles, and the Navier-Stokes equation emerges from local mass and momentum conservation in the particle ensemble. The update of particle positions and momenta mimics the underlying kinetics and is defined in terms of successive streaming and collision steps, cf.~Fig.~\ref{fig:mpc-algorithm}. During the streaming step, the particles move ballistically
\begin{equation}
\rvec_i(t+h) = \rvec_i(t) + h \rev{\vvec_i(t)}
\end{equation}
where $h$ is the time interval between collisions.
In the collision step, the particles are sorted into cubic collision cells of size $a$, and interactions occur only between particles within the same cell. The particles in each cell exchange momentum while the momentum of the collision cell is conserved
\begin{equation}\label{eq:mpc-collisions}
  \rev{\vvec_i(t+h)} = \uvec_C(t) + \bm{\Delta}_i(t)
\end{equation}
where $\rvec_C$ and $\uvec_C$ are the position and velocity of the centre of mass of the cell, respectively, and $\bm{\Delta}_i$ is a collisional term that does not change $\uvec_C$. The centre of mass velocity in MPC plays a similar role as the equilibrium distribution in LBM. 


The introduction of the collision cell introduces an artificial reference frame and breaks Galilean invariance. If the mean free path $\lambda=h\sqrt{k_BT/m}$ is smaller than the cell size $a$, repeated collisions between the same particles lead to correlations and the transport coefficients become dependent on imposed flow fields. In order to restore Galilean invariance, a random shift of the cell grid is performed before each collision step\cite{Ihle:2001,Ihle:2003a}. In practice, the shift can be performed by moving all particles by a random vector $\vec{s}$ with components uniformly distributed in $[-a/2,a/2[$ before the collision step and back after the collisions. The grid shift promotes momentum transfer between the cells and thus can lead to additional correlations in the transport coefficients\cite{Ihle:2001,Ihle:2003a}. \rev{Huang et al.\cite{Huang:2012} have carefully analysed the velocity correlations and the characteristic scales over which the correct hydrodynamic correlations and the long-time tail emerge.}


The original \rev{multi-particle collision} algorithm is referred to as stochastic rotation dynamics (SRD). It consists of a random rotation of the relative velocities $\vvec_{i,C} = \vvec_i-\uvec_C$ of the particles in a cell\cite{Malevanets:1999}
%
%
%
\begin{equation}
  \begin{multlined}
    \bm{\Delta}^\text{SR}_i(t) = \vvec_{i,C}^{\parallel}(t)
    + \vvec_{i,C}^{\perp}(t) \cos(\alpha)
    + \left(\vvec_{i,C}^{\perp}(t) \times \vec{\hat{a}}\right) \sin(\alpha)
  \end{multlined}
\end{equation}
where $\vec{\hat{a}}$ is a randomly chosen axis, $\alpha$ the rotation angle, and $\vvec_{i,C}^{\parallel}$ and $\vvec_{i,C}^{\perp}$ denote the parallel and perpendicular components of \rev{$\vvec_{i,C}$} with respect to the random axis $\vec{\hat{a}}$. This collision rule is denoted as MPC-SR$-a$\cite{Noguchi:2008}. Instead of rotating the relative particle velocities, it is also possible to dampen the velocities with a Langevin thermostat (MPC-LD) or simply generate new relative velocities randomly. The latter is implemented by the algorithm
\begin{equation}\label{eq:MPC-AT-a}
\bm{\Delta}^\text{AT}_i(t) = \vvec_i^\text{ran} - \sum_{j\in C} \frac{\vvec_j^\text{ran}}{n_C}
\end{equation}
where $\vvec_i^\text{ran}$ is a random velocity drawn from a Maxwell-Boltzmann distribution, and $n_C$ is the number of particles in the collision cell. This collision \rev{rule} is denoted as MPC-AT$-a$.\cite{Noguchi:2008} \rev{The random velocity serves as an Anderson-like thermostat to control the temperature, such that the simulation is effectively performed in a (local) canonical ensemble. For choices of alternative thermostats and their performance in equilibrium and non-equilibrium flows, we refer the reader to Refs. \citenum{Bolintineanu:2012} and \citenum{Huang:2015}.}

A drawback of both MPC-SR$-a$ and MPC-AT$-a$ algorithms is that they generate a non-symmetric stress tensor and hence do not conserve angular momentum.
This can be mitigated by imposing angular momentum conservation as a constraint, which leads to an additional term in the collision rule
\begin{equation}
\begin{multlined}
  \bm{\Delta}^{+a}_i(t) = - \rvec_{i,C}(t) \times m \tens{I}^{-1} \sum_{j\in C} \left[ \rvec_{j,C}(t) \times \left( \vvec_{j,C}(t) - \bm{\Delta}^{-a}_j(t) \right) \right]
\end{multlined}
\end{equation}
where $m$ is the particle mass, $n_C$ the number of particles in the collision cell, $\tens{I}$ the moment of inertia tensor, and $\rvec_{i,C}=\rvec_i - \rvec_C$ the relative particle position. 
%
%
MPC-LD, MPC-AT, and MPC-SR with the $\bm{\Delta}^{+a}$ term do not conserve kinetic energy. For MPC-SR, a velocity rescaling can be applied to conserve energy, however, this breaks time-reversal symmetry and leads to deviations in the radial distribution function\cite{Noguchi:2008}. A collision \rev{rule} that conserves both energy and angular momentum can be derived in two dimensions\cite{Ryder:2005}. Various other collision rules have been proposed in the literature, and we refer the interested reader to the overview by Gompper et al.\cite{Gompper:2009}

\subsection{Transport Coefficients}

\begin{table*}
\caption{Correlation factors for MPC collision algorithms. Note that the MPC$+a$ expressions for $s$, $c$ and $g$ are approximations for large $N$. For small $N$, additional correction terms have to be taken into account \cite{Noguchi:2008}.
The factor $f=\frac{N-1+\exp(-N)}{N}$ is obtained by averaging over a Poisson distribution for the cell occupation number $n_C$ with $\langle n_C \rangle = N$. For angular momentum conserving collision rules, additional correlations are present which result in the more complicated expressions. It should be noted that in the expression for $\theta$ terms of $O(1/M^2)$ have been neglected \cite{Tuezel:2003}.}
\centering
\begin{tabular}{llllll}
\hline
correlation factor & & MPC-SR & MPC-LD & MPC-AT$-a$ & MPC-AT$+a$ \\
\hline
$s$ & & \multirow{2}{*}{$\frac{2}{d} (1-\cos\alpha) f$}
      & \multirow{2}{*}{$\frac{\gamma h/m}{1+\gamma h/2m}$}
      & \multirow{2}{*}{$f$}
      & \multirow{2}{*}{$1-\frac{d+1}{2N}$} \\\\
$c$ & $d=2$ & $(1-\cos2\alpha) f$
      & \multirow{2}{*}{$\frac{2\gamma h/m}{(1+\gamma h/2m)^2}$}
      & \multirow{2}{*}{$f$}
      & \multirow{2}{*}{$\frac{d}{2N}+1\left(1-\frac{3d+2}{4N}\right)$} \\
    & $d=3$ & $\frac{2}{5} (2-\cos\alpha-\cos2\alpha) f$ \\
$g$ & & \multirow{2}{*}{$\frac{2}{d} (1-\cos\alpha) f$}
      & \multirow{2}{*}{$\frac{\gamma h/m}{1+\gamma h/2m}$}
      & \multirow{2}{*}{$f$}
      & \multirow{2}{*}{$\frac{1}{2}\left(1 - \frac{7}{5 N}\right)$} \\\\
$\theta$ & & $\frac{2}{d} (1-\cos\alpha)$ \\
& & $\quad - \frac{4}{dN} (1-\cos\alpha)^2 \left( \frac{7-d}{5} - \frac{1}{4} \csc^2\frac{\alpha}{2} \right)$ \\
\hline
\end{tabular}
\label{tab:mpc-correlation-factors}
\end{table*}

Local hydrodynamic fields are defined for each MPC cell as
\begin{align}
\rho(\xvec_C) &= \frac{m}{a^3} \sum_{i\in C} 1 , &
\jvec(\xvec_C) &= \frac{m}{a^3} \sum_{i \in C} \vvec_i , &
e(\xvec_C) &= \frac{m}{a^3} \sum_{i \in C} \frac{\vvec_i^2}{2} .
\end{align}
These coarse-grained fields are the slow variables of the discrete-time dynamics that exhibit hydrodynamic behaviour on macroscopic time and length scales.
The corresponding transport coefficients emerge from the micro-scale transport during streaming and collisions and hence contain both kinetic and collisional contributions. There are two possible routes to derive transport coefficients of the MPC fluid. The first uses a projection-operator formalism and relates the transport coefficients to equilibrium fluctuations of the hydrodynamic fields \cite{Ihle:2003b,Ihle:2005,Tuezel:2003}. In the second approach, transport coefficients are determined through analysis of non-equilibrium steady-state situations. By virtue of the fluctuation-dissipation theorem, the linear response of the system to imposed gradients allows to calculate transport coefficients that are identical to the ones obtained from equilibrium fluctuations \cite{Kikuchi:2003,Pooley:2005,Noguchi:2008}.

The latter approach leads to generic expressions for the diffusion coefficient,
\begin{equation}
D = \frac{k_BT h}{m} \left( \frac{1}{s} - \frac{1}{2} \right),
\end{equation}
and the kinetic and collisional contributions to the shear viscosity,
\begin{align}
\nu^\text{kin} &= \frac{k_BT h}{m} \left( \frac{1}{c} - \frac{1}{2} \right) ,\\
\nu^\text{col} &= \frac{g}{12}\frac{a^2}{h} ,
\end{align}
where the correlation factors $s$, $c$ and $g$ depend on the specific collision \rev{rule} and are given in Table \ref{tab:mpc-correlation-factors}. Other transport coefficients can be derived along similar lines. In addition to mass and momentum transfer, the energy conserving versions of MPC are able to reproduce the heat transfer in the fluid. The thermal diffusivity and the thermal conductivity are given by\cite{Ihle:2005,Pooley:2005}

\begin{align}
D_T^\text{kin} &= \frac{k_BTh}{m} \left( \frac{1}{\theta} - \frac{1}{2} \right), \\
D_T^\text{col} &= \frac{a^2}{3(d+2)h} \frac{1-\cos\alpha}{N} .
\end{align}

A summary of the results is reported in Table \ref{tab:mpc-correlation-factors}. The similarity of the kinetic viscosity with the corresponding expression for the lattice Boltzmann method (LBM) should be noted. The latter can indeed be derived with the approach used in this section, which shows that the kinetic viscosity of both MPC and LBM is that of an ideal gas. The LBM does not have an analogue of the collisional viscosity, however, because the LBM collision process is entirely local and does not transfer momentum. This is an essential difference between particle-based and fully kinetic methods, and here MPC can be seen as a bridge between these two approaches.


\subsection{Boundary Conditions and Suspended Objects}

\label{sec:coupling}

If the fluid is bounded by external walls or contains solid obstacles, no-slip boundary conditions at the solid surface are commonly imposed. A standard mesoscale procedure is to apply the bounce-back rule where the velocity $\vec{v}$ of an impinging particle is reversed to $-\vec{v}$. This procedure can be used if the boundaries coincide with the boundaries of the collision cells. However, due to the grid shift this is not always the case, and the boundaries generally intersect collision cells which can lead to an artificial slip velocity. This consequence of partially filled cells can be mitigated by filling the intersected cells with virtual particles such that the total number of particles matches the average of $M$ particles per cell. The MPC collision step is performed using the mean velocity of all particles in the cell, where the velocities of the virtual particles are drawn from a Maxwell-Boltzmann distribution with zero mean velocity \cite{Lamura:2001}.
%
%
The virtual particles ensure that a no-slip boundary condition is obtained which has been verified numerically for Poiseuille flows and flows around circular and square cylinders \cite{Lamura:2001}. \rev{For cases where the boundary location does not coincide with the average center of mass of the particles, artificial slip can be reduced by allowing fluctuations in the number particles number such that the number of particles is Poisson distributed\cite{Winkler:2009,Bolintineanu:2012,Huang:2015}.}


A similar approach can be used to impose boundary conditions on suspended objects such as colloidal particles. A fluid particle that collides with the colloid acquires a new velocity where the normal and tangential components are drawn from Maxwellian-like distributions \cite{Hecht:2005}.
In dense suspensions, there may be multiple collisions with more than one colloid (or a wall) during a time step, and repeated scattering is necessary to avoid an artificial depletion interaction between the colloids. The thermal wall approach is effective for suspended objects that are much larger than the mean free path $\lambda$.

For suspended objects with internal degrees of freedom, such as polymers or cells, a different coupling approach can be used. The particles are typically updated by a molecular dynamics scheme that takes into account intra- and inter-molecular forces. For the coupling to the fluid, the particles or monomers are considered point-like and interact with the fluid particles by means of the MPC collisions.
The momentum exchange $\Delta \vec{p}$ between the fluid particles and the monomers can be included as force in the molecular dynamics update. In order to accurately capture the hydrodynamic interactions, the average number of monomers per collision cells should be smaller than $1$, and the monomers should be neutrally buoyant, i.e, the monomer mass $m_\text{m}$ on the order of solvent mass per cell $n_c m$. Typically, the size $a$ of the MPC cells has to be on the order of the bond length of the polymer. The MPC point-particle coupling has been used extensively to simulate polymeric and colloidal suspensions \cite{Gompper:2009}.

\subsection{Non-Ideal Fluids, Binary Mixtures, Viscoelastic Fluids}


In order to model dense gases and non-ideal liquids, excluded volume effects have to be incorporated. Generally, any non-ideal effect makes it necessary to consider non-local interactions. This can be achieved by dividing a collision cell into sub-cells with side length $a/2$ which exchange momentum during the collision process \cite{Ihle:2006,Ihle:2008}.
%
%
The equation of state of the non-ideal MPC fluid can be found from the mechanical definition of pressure. 
The non-ideal algorithm does not conserve phase-space volume because the collision probability depends on the difference of sub-cell velocities which can be realised by different states. However, careful consistency checks have not revealed a violation of detailed balance or other inconsistencies \cite{Ihle:2008}.


Repulsive interactions between different species of a binary mixture can be achieved in a similar manner as for excluded volume interaction between sub-cells. However, instead of exchanging momentum between the entire sub-cells, only collisions between particles of type A and B are taken into account. 
Additional MPC collisions at the cell level are incorporated for momentum exchange between particles of the same type $A$ or $B$ which allows to tune the overall viscosity \cite{Tuezel:2006,Ihle:2006,Ihle:2008}.
The mixture model can be extended to amphiphilic and ternary fluids and allows to study the phase behaviour of complex fluids on large time scales.
An alternative approach to binary mixtures is the introduction of a colour charge, similar to colour models in the LBM. The collision process takes into account the concentration of the species by exchanging momentum such that the colour-weighted momentum in each collision cell points in the direction of colour gradient. While it can be used in two and three dimensions, the colour model does not include thermal fluctuations of the order parameter. Nonetheless, it has been shown that his model leads to phase separation, satisfies the Laplace equation and can be extended to simulate ternary mixtures \cite{Hashimoto:2000,Sakai:2000,Inoue:2004,Inoue:2006}.


Viscoelastic behaviour can be modelled within MPC by using dumbbell springs instead of point particles. The MPC algorithm can still be performed in the usual manner, where in the streaming step the centre of mass of the dumbbells moves ballistically while the relative coordinates
are updated according to the bond interaction.
The collision step is applied to the end-points of the dumbbells and proceeds in the usual way for the various collision \rev{rules}.
%
An MPC fluid consisting of harmonic dumbbells can capture the orientation and elongation in shear flow and thus reproduces the viscoelastic behaviour of a Maxwell fluid \cite{Tao:2008}. However, due to the possible infinite bond extension, harmonic dumbbells do not reproduce non-equilibrium properties such as shear-thinning. If FENE dumbbells are employed instead, the proper behaviour corresponding to a dilute polymer solution is found \cite{Ji:2011}. Another alternative is to introduce a constraint of constant mean square bond length where the equilibrium value corresponds to Gaussian chains \cite{Kowalik:2013}.
%
%
The dumbbell fluid exhibits shear thinning, where at large shear rates $\eta_b \sim {Wi}^{-2/3}$ with the Weissenberg number ${Wi}=\dot{\gamma} k_B T/(2DK_0)$. Moreover, due to hydrodynamic interactions, the fluid exhibits a non-vanishing second normal stress with the same shear rate dependence \cite{Kowalik:2013}.

\subsection{Applications}




Like in other mesoscopic methods, MPC has been applied to a wide range of colloidal and polymeric systems \cite{Winkler:2004,Winkler:2005,Gompper:2009}.
The capabilities of MPC for simulating complex fluids are exemplary highlighted for the hydrodynamics of star polymers in shear flow \cite{Ripoll:2006,Ripoll:2007,Singh:2014}. Star polymers are modelled as linear polymer chains that are linked to a common monomer at the centre.
A shear flow is induced in the MPC solvent by employing Lees-Edwards boundary conditions.
The imposed flow field is strongly screened inside the star polymer, which was confirmed by comparing simulations with and without hydrodynamic interactions, respectively \cite{Ripoll:2007}.
The results from MPC simulations were found to agree well with experimental measurements \cite{Singh:2014}.

\begin{figure}[h]
  \centering
  \includegraphics[height=.23\textheight]{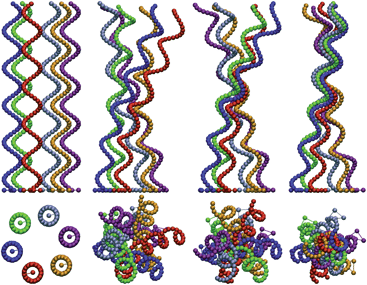}
  \caption{Synchronisation and bundling of helical flagella are governed by hydrodynamic interactions and can be modelled using an MPC fluid. Reproduced from Ref. \citenum{Reigh:2012} with permission from the Royal Society of Chemistry.}
  \label{fig:ecoli-swimmer}
\end{figure}


MPC can also be applied to non-equilibrium flows where analytical theories are still limited due to the lack of applicable variational principles. One particular example of a driven dissipative system are microfluidic droplets in a Hele-Shaw geometry. 
A train of microfluidic droplets can be modelled within 2D-MPC using a frictional coupling on disc-like droplets \cite{Fleury:2014,Schiller:2015}.
The results confirm quantitatively that a far-field approximation of the dipolar hydrodynamic interactions remains valid even at high densities \cite{Fleury:2014}. Moreover, a confinement-induced coupling of longitudinal and transverse oscillations was discovered where the longitudinal motion of the droplets is induced by the boundary conditions of the flow field \cite{Fleury:2014,Schiller:2015}. In the presence of thermal fluctuations, the droplet oscillations exhibit instabilities that were also observed in simulations \cite{Schiller:2015}. MPC thus paves the way to systematic investigation of the governing principles of non-equilibrium systems using well controlled model systems that are at the same time promising to develop novel methods in statistical physics.


Another area of high interest where timely contributions have been made using MPC is the hydrodynamics of swimmers \cite{Lauga:2009,Elgeti:2015}.
Swimming behaviour can be produced by a number of mechanisms which differ in the characteristics of the flow field they produce and, consequently, the governing hydrodynamic interactions. For example, peritrichous bacteria use rotating helical flagella for self-propulsion, cf.~Fig.~\ref{fig:ecoli-swimmer}.
The mechanism of synchronisation and bundling of flagella has been studied using MPC \cite{Reigh:2012,Reigh:2013,Hu:2015}. The dependence of characteristic times for synchronisation and bundling on the number of flagella, separation, and motor torque was found to be governed by hydrodynamic interactions.
%
The swimming properties of a flagellar model bacterium consisting of a spherocylindrical body have also been simulated using MPC \cite{Hu:2015}.
These simulation models can be extended to study swimming near surfaces where confinement effects and non-equilibrium effects are expected to play an important role, similar to the microfluidic droplet systems described above.


\begin{figure}[h]
  \centering
  \includegraphics[height=.23\textheight]{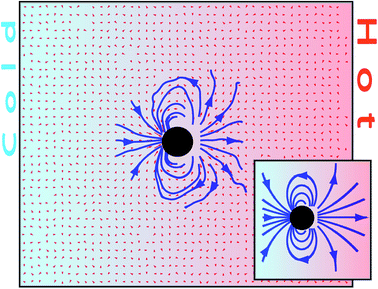}
  \caption{Flow field around  a freely drifting colloidal sphere in a temperature gradient. Reproduced from Ref. \citenum{Yang:2013} with permission from the Royal Society of Chemistry.}
  \label{fig:colloid-thermophoresis}
\end{figure}

MPC is uniquely suited to study inhomogeneous systems with temperature gradients. The investigation of thermodiffusion and thermophoresis using MPC was pioneered by Ripoll and co-workers \cite{Yang:2011,Luesebrink:2012a,Yang:2013,Yang:2014a,Yang:2014b,Yang:2014c,Yang:2015}.
The flow field induced by thermophoretic motion of a colloidal particle in an MPC fluid was found to be Stokeslet-like for a fixed particle and dipolar for a freely drifting particle, cf.~Fig.~\ref{fig:colloid-thermophoresis} \cite{Yang:2013}.
The study of thermophoretic motion using MPC has inspired a number of applications, for instance, the induced flow field around a fixed colloidal particle gives rise to a net solvent flow which can be exploited as a microfluidic pump \cite{Yang:2013}. Another interesting application is the case where the temperature gradient is generated locally by a solute particle, e.g., by inhomogeneous light absorption or inhomogeneous chemical reactions. This idea was subsequently used to design a thermophoretic nanoswimmer consisting of two linked monomers and self-phoretic Janus particles \cite{Yang:2011,Yang:2014a}.
Rotational motion due to thermophoresis was first demonstrated in MPC simulations of a micro-gear \cite{Yang:2014b}, and subsequently for a micro-turbine with anisotropic blades\cite{Yang:2014c} and a catalytic micro-rotor driven by diffusiophoresis in a concentration gradient \cite{Yang:2015}.


\section{Lattice Boltzmann Method}

The lattice Boltzmann Method (LBM) \cite{Chen:1998, Succi:2001, Aidun:2010, Guo:2013, kruger2016lattice} is a lattice-based model unlike the particle-based methods described in the previous sections. Historically, it has not evolved from an adaptation of a flavour of molecular dynamics, but from lattice gas cellular automata \cite{frisch_lattice-gas_1986}. LBM was originally introduced by McNamara and Zanetti \cite{McNamara:1988} to mitigate the statistical noise that plagued the lattice gas automata. LBM is rooted in the kinetic theory of gases, and solves a simplified and discretised Boltzmann equation.

Characteristic for LBM is the full discretisation of time, configuration and velocity space\cite{shan_kinetic_2006}. The configuration and velocity discretisation are matched in such a way that the resulting algorithm becomes favourably simple and nearly local. These features lead to high numerical efficiency and parallelisability which are today extensively exploited throughout the soft matter community. 

\subsection{Basic Algorithm}

\begin{figure}[h]
 \centering
 \includegraphics[width=.9\columnwidth]{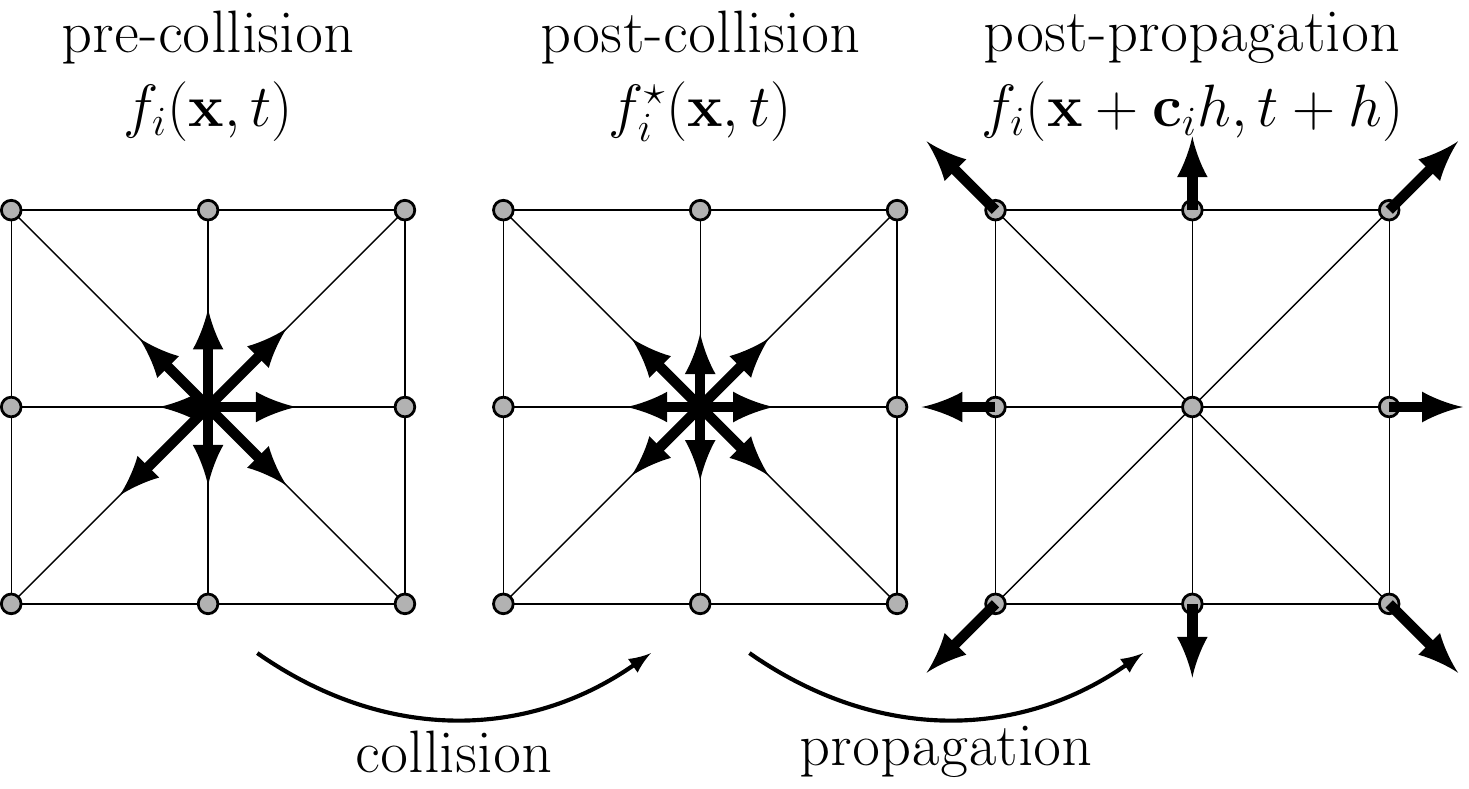}
 \caption{The lattice Boltzmann equation, Eq.~(\ref{lbe}), can be understood as subsequent local collision and non-local propagation steps. Each arrow indicates the magnitude and direction of one of the nine distributions $f_i$ on a 2D lattice during each of the stages of the algorithm. Lattice nodes are shown as grey circles.}
 \label{fig:lbm-algorithm}
\end{figure}

The central quantity for LBM is the probability density function
$f(\mathbf{x}, \mathbf{c}, t)$ which measures the probability of finding
a particle with velocity $\mathbf{c}$ at position $\mathbf{x}$ at time $t$.
Under the assumption of molecular chaos, $f$ is a one-particle distribution function, i.e., the velocities of all colliding particles are uncorrelated and independent of their positions.
Molecular chaos implies that the Boltzmann equation can be closed, i.e., the collision terms does not include the two-particle distribution function.
This assumption is well justified for gases and fluids which are not too dilute (low Knudsen number).
Then $f$ characterises the dynamic state and obeys the Boltzmann equation.
The basic algorithm of its simplified and discrete analogue, the lattice Boltzmann equation (LBE),
is shown in Fig.~\ref{fig:lbm-algorithm} and can be summarised as
\beq\label{lbe}
f_i(\mathbf{x}+\mathbf{c}_i\, h, t+h)=f_i^\star(\mathbf{x},t)=f_i(\mathbf{x}, t)+\Omega_i(\mathbf{x},t) + S_i(\mathbf{x}, t)
\eeq
with $\Omega_i$ and $S_i$ explained below.
The velocities $\mathbf{c}_i$ are a set of discrete lattice vectors that
connect each lattice site to its nearest neighbours. 
Due to the discrete nature of the lattice vectors, the isotropy is generally broken. 
However, through an appropriate choice of the $\mathbf{c}_i$, isotropy can be restored to the required extent, and the resulting $n$-dimensional LBM schemes with $m$ velocities are often denoted D$n$Q$m$ in the popular classification according to Qian\cite{Qian:1992}.

The term $\Omega_i$ in Eq.~(\ref{lbe}) is the collision operator that describes the effective 
intermolecular interactions that modify the distribution function $f_i$ related to each
lattice direction. It can be written in the general form
\begin{equation}\label{mrt_col_op}
\Omega_i
= - \sum_j \Lambda_{ij} \left( f_j - f_j^\text{eq} \right)
= - w_i \sum_k b_k^{-1} e_{ki} \lambda_k \left( m_k - m_k^\text{eq} \right)
\end{equation}
where the eigenvalues $\lambda_k$ are relaxation coefficients and the moments $m_k$ are obtained from the orthogonal transformation
\begin{align}
m_k &= \sum_i e_{ki} f_i , & f_i &= w_i \sum_k b_k^{-1} e_{ki} m_k
\end{align}
and $e_{ki}$ are orthogonal basis vectors with respect to the scalar product $\sum_i w_i e_{ki}e_{li}=b_k\delta_{kl}$. The basis vectors $e_{ki}$, weights $w_i$, and norms $b_k$ depend on the specific lattice model in use~\cite{dHumieres:2002,Duenweg:2009}.
This model is called multiple relaxation times (MRT).
The local equilibrium values $m_k^\text{eq} = \sum_i e_{ki} f_i^\text{eq}$ are chosen such that the macroscopic equations are correctly recovered and are often based on a low-velocity expansion of the Maxwell-Boltzmann distribution:
\beq
\label{LBM_eq}
f_i^\text{eq}(\rho, \mathbf{u}) = w_i\,\rho\left(1 + \frac{\mathbf{u}\cdot\mathbf{c}_i}{c_\text{s}^2} + \frac{\vec{u}\vec{u}:(\vec{c}_i\vec{c}_i-c_\text{s}^2\tens{I})}{2c_\text{s}^4} \right).
\eeq

The simplest choice of the eigenvalues is the single relaxation time approximation $\lambda_k=\tau^{-1}$, known as the lattice-BGK collision model named after Bhatnagar, Gross and Krook (BGK), or single relaxation time (SRT). The relaxation time $\tau$ is related to the kinematic viscosity $\nu$ via 
$\nu = (\tau - h/2)\,c_\text{s}^2$ with the speed of sound $c_\text{s}$.
The MRT model offers a larger number of free parameters than BGK; these parameters can be used to increase accuracy and stability of the algorithm.
MRT also allows to have different shear and bulk viscosities.
Other collision operators are the so-called two relaxation times (TRT)\cite{ginzburg_TRT_2008} or entropic LBM\cite{Ansumali:2003}.

The source term $S_i$ in Eq.~(\ref{lbe}) is related to the local volumetric force $\vec{F}$ through \cite{Guo:2002}
\beq
\label{LBM_Guo}
S_i = \sum_j \left( \delta_{ij} - \frac{1}{2} \Lambda_{ij} \right) w_j \left(\frac{\vec{F}\cdot\vec{c}_j}{c_\text{s}^2} + \frac{\vec{u}\vec{F}:(\vec{c}_j \vec{c}_j - c_\text{s}^2 \tens{I})}{2c_\text{s}^4}\right) .
\eeq

The density $\rho$, momentum density $\rho \mathbf{u}$, and momentum flux $\bm{\Pi}$ of the fluid are the zeroth, first, and second moments, respectively~\cite{Duenweg:2009, kruger2016lattice}:
\begin{align}
\rho &= \sum_i \, f_i, & \rho \mathbf{u} &= \sum_i \, \mathbf{c}_i f_i + \frac{\vec{F}}{2}, & \bm{\Pi} &= \frac{1}{2} \sum_i \left( f_i+f_i^\star \right)\vec{c}_i\vec{c}_i .
\end{align}
The momentum flux $\bm{\Pi}=\bm{\Sigma}+\bm{\sigma}$ contains the Euler stress $\bm{\Sigma} = \rho \mathbf{u} \mathbf{u} + p \mathbf{I}$ and the deviatoric stress tensor $\bm{\sigma} = \rho \nu [\bm{\nabla} \mathbf{u} + (\bm{\nabla} \mathbf{u})^\top]$.
The link between the LBE (Eq.~\ref{lbe}) and the macroscopic Navier-Stokes equation can be established through a Chapman-Enskog expansion \cite{Succi:2001, Duenweg:2009, kruger2016lattice}.

A direct consequence of the discreteness of the underlying lattice is that Galilean invariance is broken in the LBM, which manifests itself as $O(u^2)$ errors in the viscosity. Hence the LBM is valid in the incompressible limit which corresponds to low Mach numbers. It is also possible to devise mitigating schemes \cite{Chikatamarla:2006, Dellar:2014}. They tend, however, to take away some of LBM's simplicity.

In the standard LBM algorithm, hydrodynamic flow occurs without thermal fluctuations. 
A scheme of fluctuating LBM can be devised which satisfies a
fluctuation-dissipation theorem \cite{Adhikari:2005}. The thermalisation of
the so-called ghost modes, the additional degrees of freedom that are not directly related to the 
ten hydrodynamic observables, lead to equipartition of the fluctuating energy on all length scales. 
This scheme has been recently extended from single phase fluids to binary mixtures 
\cite{Gross:2010}.

\subsection{Boundary Conditions}

The LBM knows countless boundary conditions \cite{kruger2016lattice}.
The reason for this plethora is that there are no unique links between the macroscopic Dirichlet and Neumann boundary conditions and the boundary conditions for the kinetic quantities $f_i$.
Here, we will briefly mention only the most common boundary conditions for soft matter applications.

The bounce-back (BB) boundary condition\cite{cornubert_knudsen_1991} takes direct advantage of the kinetic nature of LBM.
As such, there is no counterpart in conventional Navier-Stokes solvers, showing the advantage of LBM for applications with complex geometries.
A post-collision population $f^\star_i$ initially moving from a fluid site $\mathbf{x}$ to a solid site $\mathbf{x} + \mathbf{c}_i h$ reverts its direction (denoted by $\mathbf{c}_{\bar{i}} = - \mathbf{c}_i$) during the BB procedure, only to reach its origin at the end of the propagation step:
\begin{equation}
 \label{lbm_bb}
 f_{\bar{i}}(\mathbf{x}, t + h) = f^\star_i(\mathbf{x}, t) - 2 w_i \rho\frac{\mathbf{c}_i \cdot \mathbf{u}_\text{b}}{c_\text{s}^2}
\end{equation}
where the second term on the right-hand side is the momentum exchange due to the boundary moving with velocity $\mathbf{u}_\text{b}$. 
Eq.~(\ref{lbm_bb}) replaces the LBE in Eq.~(\ref{lbe}) if the node at $\mathbf{x} + \mathbf{c}_i h$ is solid rather than fluid.

BB leads to a no-slip boundary condition at a plane half-way between the fluid and the solid nodes with second-order accuracy\cite{ginzbourg_boundary_1994}.
Despite its staircase-like discretisation that can lead to numerical artefacts, the BB method is stable, simple to implement and computationally efficient. It is often employed for colloidal particles and porous media\cite{ladd_numerical_1994, aidun_lattice_1995}.
More sophisticated boundary conditions exist that allow boundaries at other locations than half-way between lattice nodes, for example interpolated BB\cite{bouzidi_momentum_2001} or multireflection\cite{ginzburg_multireflection_2003}.
We refer to a comprehensive review \cite{Duenweg:2009} for further information on fluid-structure interactions for soft matter applications.

In order to model immersed soft objects, such as polymers\cite{Ahlrichs:1999} and biological cells\cite{bagchi_mesoscale_2007}, the immersed boundary method (IBM)\cite{peskin_immersed_2002} is commonly used.
The basic idea of the IBM is to introduce a Lagrangian mesh that captures the shape of the immersed object and can move relatively to the Eulerian lattice of the fluid.
The mesh and the lattice are coupled through velocity interpolation (Eulerian to Lagrangian) and force spreading (Lagrangian to Eulerian) steps.
It is possible to equip the Lagrangian objects with suitable elastic properties, such as dilation, shear and bending resistance.

\rev{Several works have been published about the combination of thermal fluctuations and immersed objects, such as colloids or polymers.
Ahlrichs and D\"unweg\cite{Ahlrichs:1999} added thermal fluctuations to the particle phase and restored momentum conservation by including opposite fluid forces.
Contrarily, Ollila et al.\cite{ollila2013hydrodynamic} claim that the addition of Langevin noise to the particle phase can be avoided, an idea that is closer to Einstein's original notion of Brownian noise.
There does not seem to be a consensus under which conditions both approaches are equivalent or unphysical.}

\subsection{Multi-Phase and Multi-Component Flows}

Multi-phase or multi-component flows are ubiquitous in soft matter systems and 
occur generally in mixtures of different species, heterogeneous systems like fluid-solid or fluid-gas mixtures 
or when interfaces are present. LBM offers a convenient route to describe the forces  
arising between the individual phases and components in a consistent way, and probably even more importantly, 
on the mesoscopic level and in complex geometries.

Several methods have emerged to date. The Shan-Chen (SC) model \cite{Shan:1993}, for instance, is a 
phenomenological approach which is based on so-called pseudo-potentials that mimic the microscopic 
interactions between the constituents.
For a fluid mixture with an arbitrary number of components, each component $\sigma$ is subjected to the short-range interaction force
\begin{equation}
 \mathbf{F}^\sigma(\mathbf{x}) = -\psi^\sigma(\mathbf{x}) \sum_{\sigma'} g_{\sigma \sigma'} \sum_{i} w_i \psi^{\sigma'}(\mathbf{x} + \mathbf{c}_i h) \mathbf{c}_i h
\end{equation}
where $g_{\sigma \sigma'} = g_{\sigma' \sigma}$ is the interaction strength between components $\sigma$ and $\sigma'$.
Note that $\sigma = \sigma'$ is permitted, which allows for self-interaction and non-ideal fluids with phase change.
The pseudo-potential $\psi^\sigma(\mathbf{x})$ is a (usually linear or exponential) function of the component density $\rho^\sigma(\mathbf{x})$ and has been introduced to increase the numerical stability of the method.
Surface tension emerges from the component interaction; its value $\gamma$ is a function of the interaction strength and the chosen pseudo-potential.

An alternative approach is based on free energy (FE) models \cite{Swift:1995, Swift:1996}.
In contrast to the SC method with its emergent surface tension properties, FE methods are based on a top-down approach, starting directly from the relevant 
target equation that describes the dynamics of the complex fluid.
Therefore, opposed to the SC model, thermodynamic consistency is straightforwardly realised in 
FE models. This is because the latter use the free energy functional as input,
which means thermodynamic coupling terms between order parameters can be directly specified.
Hence, FE models can be systematically extended to incorporate additional physical effects.
They also provide better control of the thermodynamic state, an important aspect close to 
interfaces and boundaries.
For example, the Landau free energy density for a binary mixture of two immiscible liquids 
with identical density $\rho$ for both liquids can be written as\cite{briant_lattice_2004}
\beq
\label{lbm_freeenergy}
\psi = c_\text{s}^2 \rho \ln \rho + \frac{a}{2} \phi^2 + \frac{b}{4} \phi^4 + \frac{\kappa}{2} (\nabla \phi)^2
\eeq
where $\phi = \frac{\rho_A -\rho_B}{\rho_A +\rho_B}$ is the local order parameter (i.e. $\phi=+1$ 
for liquid $A$ and $\phi=-1$ for liquid $B$) 
and $a, b$ and $\kappa$ are free parameters.
Phase separation can occur when $a < 0$, and the surface tension is given by $\gamma = \sqrt{8 \kappa a^3 / 9 b^2}$.
The chemical potential is the functional derivative of the free energy:
\beq
 \mu = \frac{\delta \psi}{\delta \phi} = a \phi + b \phi^3 - \kappa \nabla^2 \phi.
\eeq

More recently, hybrid approaches have emerged \cite{Marenduzzo:2007}. 
\rev{In order to understand their novelty,
it is necessary to realise that any additional equation of motion beside the Navier-Stokes equation
that is involved in the dynamics of the complex fluid can be solved
as well with an LBM-style method by defining additional sets of 
either scalar, vectorial or tensorial distribution functions.
The corresponding observables are then obtained by taking moments
of these distribution function is just the same fashion.
The equilibrium distributions are similar to Eq. \ref{LBM_eq}, but have 
different coefficients owing to the different target equation. 
A good example of this full-LB approach has been given for the dynamics of 
liquid crystals \cite{Denniston:2004}, where a partial differential equation
for the tensorial order parameter, often referred to as Q-tensor, needs to be solved on 
top of the Navier-Stokes equation. Full-LB approaches can feature increased stability compared to 
alternative methods. But on the down side they require more memory 
and floating point operations per timestep.}

\rev{Hybrid schemes do not define additional sets of distribution functions and use 
LBM only for the Navier-Stokes part of the problem. Any partial differential equation
is solved with finite-difference schemes. In the above example of liquid crystals and 
based on a D3Q19 LB model, this reduces the memory requirements by 
almost 79\%: Instead of 19 distributions for the Navier-Stokes equation and 5$\times$19 
distributions for the 5 independent components of the Q-tensor in a full-LB approach,
a hybrid model uses only 19 distributions for Navier-Stokes and holds the 5 independent 
components of the Q-tensor directly rather than reconstructing them from moments of the 
distribution. This, of course, is an extreme example due to the tensorial nature of the 
order parameter.}

Like other models that involve phase boundaries or interfaces, both SC and FE models suffer from spurious 
currents in regions of large order parameter gradients. They can be reduced by higher-order 
isotropic schemes \cite{Connington:2012}. We refer to the relevant literature 
\cite{Guo:2013, kruger2016lattice} for more details on multi-phase and multi-component mixtures.

\subsection{Applications}

Among the first soft matter applications of LBM were flows in porous media \cite{Chen:1998}, 
particulate flows \cite{Ladd:2001} and flows of polymer solvent systems \cite{Ahlrichs:1999}. 
These systems are notoriously difficult to model with standard CFD methods.
On the contrary, LBM offers a comparably simple way of handling boundary conditions 
and fluid-solid interactions and seems almost ideally suited. 
For instance, the flow through a porous rock material, traditionally only describable 
through effective laws such as Darcy's law, can be simulated directly.

\begin{figure}[h]
 \centering
 \includegraphics[width=.9\columnwidth]{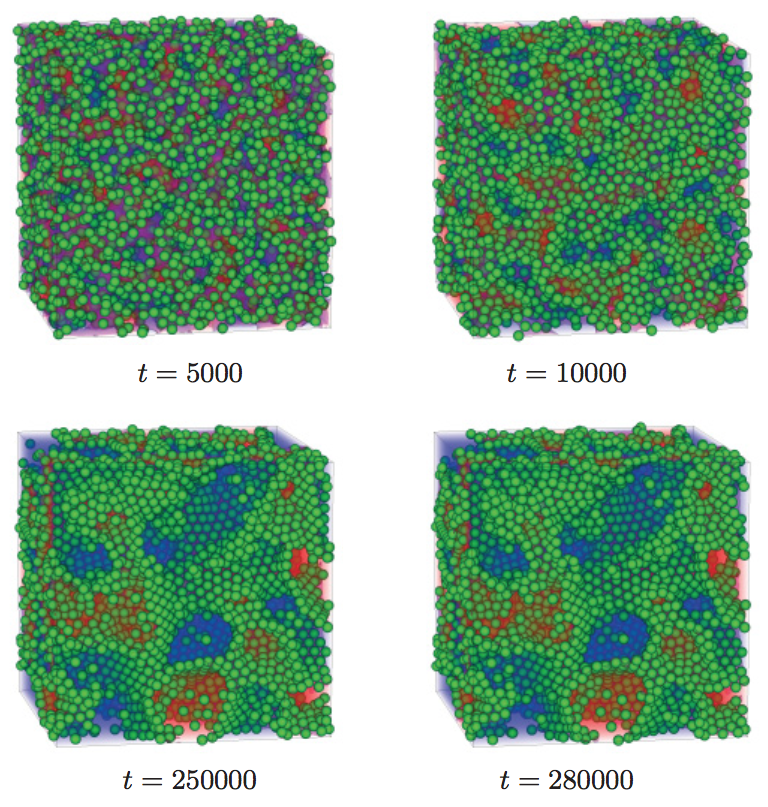}
 \caption{3D visualisation of the a bicontinuous interfacially jammed emulsion gel (bijel). Shown are the particles 
and the two fluids, respectively. The two pictures at the bottom depict the bicontinuouity of the fluids and the 
attachment of the particles to the interface \cite{Jansen:2011}.  
Reprinted figure with permission from Physical Review E. Copyright (2011) by the American Physical Society.}
\end{figure}

With regard to multi-component systems like binary \cite{Swift:1996} or amphiphilic mixtures \cite{Chen:2000}
of immiscible fluids, LBM has
as well a number of advantages. The modelling of droplet coalescence or 
necking phenomena usually requires expensive interface tracking algorithms that can
trigger numerical instabilities. In LBM, 
these systems can be modelled with two sets of 
LB distribution functions, or with one set of distributions and 
a level set in hybrid methods. 
The position of the interface is then simply defined 
by a value of the level set or by the isosurface where the densities of both fluids have the same value. 
These models have also been successfully applied to nanoparticle-stabilised suspensions 
(so-called Pickering emulsions) \cite{Jansen:2011} and wetting phenomena \cite{Blow:2010}.

Free energy models and their ability to express thermodynamic forces on 
solute and solvent through gradients of 
chemical potentials come into their own when the 
composition and local structure of the complex fluid becomes increasingly intricate.
This, for example, is the case in liquid crystals \cite{Denniston:2004} where the 
anisotropic local order structure is described through a tensorial order parameter
and gives rise to spatially varying rheological properties.
Hybrid approaches in favour of full-LB schemes have been used for 
large-scale mesoscopic simulations of 3D liquid crystalline structures 
at very low shear rates \cite{Henrich:2013}. Similar concepts have been 
used to study active liquid crystals and gels \cite{Marenduzzo:2007}. 
They can be described through the same methodology with modified
free energy functionals that account for the active components
and additional forces.

Other interesting applications comprise charged soft matter systems
and electrokinetic phenomena. An important step forward was
the development of the link-flux method \cite{Capuani:2004}
by which the leakage of charge between interfaces and surfaces could
be prevented. The scheme has been used in the mesoscopic study
of the electrophoresis of charged colloids \cite{Giupponi:2011} and 
polyelectrolytes \cite{Grass:2010}.
Recently, the scheme has been improved to reduce spurious currents
\cite{Rempfer:2016}.

\section{High-Performance Computing}

The physics of many soft condensed matter systems can be investigated without 
ever considering large-scale simulations. However, high-performance computing (HPC) 
can be of particular importance, not only for simulation methodologies 
that can capture the long-range nature 
of hydrodynamic interactions --- one focus of this tutorial review. 
Soft matter generally shows a tendency to self-organise 
into characteristic mesoscopic structures that in turn determine 
its macroscopic properties. Multiscale modelling techniques, accompanied 
by the appropriate computing power, are finding increasingly 
wide-spread use in this field of research.

The so-called rise of the machines has indeed transformed HPC
from an exotic niche technology into an indispensable tool for science and engineering.
In fact, since the beginning of this millennium, the computing power of the 
number one facility listed in the bi-annually published TOP500 has increased
by more than a factor 40,000. With the advent of even more powerful 
graphics cards and many-core processors, an end to this astonishing development 
is not in sight. 

At the same time, HPC architectures are rapidly diversifying and become more complex.  
This makes it even more difficult to program and harvest their computational power
in the future. This development began about a decade ago and is directly linked to 
the breakdown of the so-called Dennard scaling. As the size and capacitance of integrated 
circuits decreased (and continues to decrease according to Moore's law), 
Dennard scaling allowed to operate them as well at lower voltages.
Hence, the power gains in terms of lower capacitance and operating voltage could 
be ``invested'' in a higher clock frequency, making processors ever faster whilst keeping
the power consumption more or less constant.
This design principle became unviable as currents in the devices have weakened and
the operating voltage in the devices cannot be further reduced. Therefore, the faster
processors can only be created through more compute cores that run calculations in 
parallel.

In this paragraph, we will glance at the latest developments in the area of accelerators and coprocessors
and describe the requirements that these increasingly heterogeneous computing architectures
pose to the programmer. Finally, we will name a few examples of production codes used 
for soft matter research.

\subsection{General Purpose Graphics Processing Units}

General purpose graphics processing units (GPGPUs), also known as 
graphics cards or GPUs, were originally designed to handle the large amount of floating-point operations
that occur in the graphics calculations of computer games. They made their first appearance 
in the world of scientific computing around 2007. Back then, Nvidia introduced its
compute unified device architecture (CUDA), an application programming interface (API) that
vastly simplified the programming of GPUs. This has so far led to a dominance of the Nvidia 
framework in the area of scientific computing with GPUs and prevented other vendor-independent standards like the 
open computing language (OpenCL) from attaining the same level of maturity.
Alternative higher-level directive-based standards like OpenACC exist and permit automatic handling  
of data management and computational offloading at the expense of slightly reduced 
performance due to reduced control over data movements. This trade-off, however, 
makes more than up for the simplicity and portability as OpenACC-enabled code 
can run as well on many-core processors.

At the time of writing, the latest generation of Nvidia's Pascal 
architecture-based Tesla P100 GPUs delivers a peak performance of around 5 TFLOPs 
(5$\times$10$^{12}$ double-precision floating point operations per second)
on 3584 CUDA cores at a power input of around 300 W, or 3--4 laptops.
To put this in perspective, the fastest supercomputer in November 2000
was the ASCI White at Lawrence Livermore National Laboratory which 
achieved about the same performance with a power input of 3 MW
(and another 3 MW for cooling, adding up to an electricity bill of roughly
\$6 million p.a.). The Tesla P100 also features 500+ GB/s 
memory bandwidth and a unified memory space to which both GPU and CPU can point. 
This alleviates the burden of having to copy data back and forth between CPU (host) and GPU (device), 
a process that uses the traditionally slowest part in a computer and came always at significant 
 costs.

However, in order to unleash the power of the GPU, algorithmic parallelism has
to be fully exposed and mapped to the architectural parallelism of the hardware 
on which the code is deployed. The so-called task or thread-level parallelism (TLP) 
that modern accelerators offer through their hundreds or thousands of 
compute cores working in parallel is only one way to expose parallelism. 
Another concept, which dates back to the early days of vector processors, 
is known as data parallelism. The basic idea behind it is that one single instruction
processes a chunk of similar data elements at the same time. 
Today, virtually all types of processors, also multicore CPUs, use advanced
vector extensions (AVX) which are extensions to the x86 instruction set. 
These are also known as SIMD (Single Instruction Multiple Data) instructions.      
Obtaining a good performance on any kind of accelerator, co-processor or many-core chip depends
increasingly on whether data parallelism can be exploited. Quite often this necessitates
a complete redesign of the memory layout and memory access pattern, 
which can be prohibitive in the case of big legacy codes.

\subsection{Intel Xeon Phi}

While GPGPUs used to have a commanding lead over alternative concepts of co- and many-core
processors, the latest edition of Intel's many integrated core (MIC) architecture,
the Xeon Phi Knights Landing (KNL) has stirred up some competition. It consists of a smaller number
of compute cores, typically 72, which can be run in a hyper-threading mode, giving in total 
up to 288 threads. The peak performance is around 3.5 TFLOPs at an energy input
comparable to Nvidia's Tesla P100. The memory bandwidth of the KNL has been significantly increased
compared to older generations; it is at about 400+ GB/s.

The KNL gets its performance from SIMD instructions on very large vector units. Hence,
without exposing data parallelism, the performance will be noticeably degraded. In fact,
data parallelism is probably even more crucial for MICs than for GPUs.
That said, obtaining a consistently good performance with these devices, GPUs or MICs coprocessors alike,
when real-world problems rather than simplistic benchmarks are considered, remains
generally challenging. Performance measurements gained with one and the same code can vary 
substantially depending on the individual scientific problem.

Nevertheless, MICs offer an unparalleled advantage over GPUs in terms of portability. 
In order to port a code to the GPU, it is often necessary to rewrite large parts.
This frequently leads to more complex functionality and much longer source code compared to its CPU
counterparts. Once the investment has been made, the performance gain can be impressive.
But the code can be only deployed on GPUs, or when CUDA has been used even only on Nvidia GPUs.

Code written in standard programming languages such as C, C++ and Fortran using the most common 
abstractions for parallel programming, i.e., message passing interface (MPI) for 
distributed memory or OpenMP and even OpenACC for shared memory architectures, 
can be compiled for the Xeon Phi coprocessor. Hence, the code can run on a variety of architectures,
ranging from multi-core CPUs in laptops over workstations, many-core processors and computing clusters to 
supercomputers with on-node co-processors. Although to obtain a good performance the same fundamental principles 
apply as for the GPU (exposing sufficient data parallelism and providing a cache-coherent memory layout 
and access pattern), the entry barrier is much lower and the portability significantly higher.
Moreover, performance improvements made for the Xeon Phi will equally benefit when the code is run 
on simple multicore CPUs and vice versa. 

\subsection{Community Codes}

Developing efficient software for HPC applications has never been a trivial task as it normally exceeds 
the resources of a typical research group. Moreover, the pool of skilled programmers who have also sufficient 
scientific experience in their area of research is limited. This makes it difficult to conduct ambitious code projects, 
even when the funding is in place! Hence, it is not surprising that community codes have emerged
which are sometimes used by thousands of researchers across the world. Here, we will only mention 
open source codes that are freely available for academic users. The list below can only be a cross section
and does not claim to be comprehensive.

A prime example of community code is the large-scale atomic/molecular 
massively parallel simulator (LAMMPS) \cite{lammps}, developed by Sandia National Laboratories, USA. 
LAMMPS is modular and relatively easy to extend, which makes it an attractive code to base individual 
research projects on. Its latest version features a vast number of models, particle types, force fields,
ensembles and integrators. 
In fact, to our knowledge LAMMPS is the only code that contains an implementation of virtually all methods
mentioned in this review. 
That said, each method comes naturally in a multitude of different flavours and modifications. The 
LAMMPS implementation contains only some of these features. The MPCD model, for instance, uses the 
stochastic rotation dynamics version, \rev{whereas the LB model is an implementation of Ollila's above
mentioned algorithm}.
The existing implementations form an excellent starting point for further development of LAMMPS, which is 
one of the reasons why new features are constantly being added to every new release.
LAMMPS has specific strengths for coarse-grained modelling, making it suitable for soft matter
research in general. The code shows very good performance up to millions of simulated particles and thousands of
cores, and the number of multi-GPU- and many-core-enabled force fields and features is continuously growing.

Another highly versatile software package for performing many-particle molecular dynamics simulations, 
with special emphasis on coarse-grained models as they are used in soft matter research is 
ESPResSo ({E}xtensible {S}imulation {P}ackage for {RES}earch on {SO}ft matter) \cite{Limbach:2006,Arnold:2013}. 
It is commonly used to simulate systems such as polymers, colloids, ferro-fluids and biological systems, 
for example DNA or lipid membranes. ESPResSo also contains a unique selection of efficient algorithms for 
treating Coulomb interactions. Recently, several grid based algorithms such as lattice Boltzmann and an electrokinetics solver have been added.

Another interesting code is HOOMD-blue \cite{hoomd}, a general purpose multi-GPU enabled molecular dynamics 
simulation toolkit from the University of Michigan, USA. Besides Brownian, Langevin dynamics, DPD 
and a number of ensembles, the hard particle capabilities and a variety of shape classes for
Monte Carlo simulations are particularly worth mentioning.
Other important community codes are GROMACS \cite{gromacs} and NAMD \cite{namd} which are popular in the biomolecular community. 
They are both highly optimised and perform exceptionally well in term of parallel efficiency, but lack the general 
extensibility that for instance LAMMPS offers.
 
Excellent parallel performance is perhaps not trivial, but much easier to achieve for lattice Boltzmann codes.
The algorithm requires only communication between nearest neighbours on a regular grid and is intrinsically parallel ---
unlike MD-type algorithms, where only the introduction of cutoffs, neighbour lists and sophisticated 
communication patterns renders the problem parallelisable.
Nevertheless, a true LBM community code has so far not emerged.
However, both OpenLB \cite{openlb}, developed at the Karlsruhe Institute of Technology, 
Germany, and Palabos \cite{palabos}, the open source project developed in a collaboration between the University of 
Geneva and FlowKit Ltd.~Lausanne, Switzerland, come probably closest to this endeavour.
waLBerla \cite{walberla}, developed by the University of Erlangen-N\"urnberg, is another LBM code with 
particular strengths in solving partial differential equations that are coupled to the fluid flow.
\rev{Another open source LBM code specifically for soft matter is Ludwig \cite{ludwig}. 
It is being developed in a collaboration between different European project partners in the UK, 
France, Switzerland and Spain and features a wide range of mesoscopic models for complex fluids and
active matter.} 


\section{Conclusions and Outlook}

This tutorial review gives an introduction to mesoscopic modelling and
simulation of soft matter. We have focused on what we think are the
three most popular methods and have reviewed the basic algorithm and
features that are particularly relevant to mesoscale modelling. We have
highlighted the specific strengths of each method with a few selected
examples.

As noted in the introduction, the mesoscopic methods are based on
conservation of mass, momentum and energy. They differ in the way the
conservation laws are translated into a coarse-grained representation,
resulting in different degrees of freedom and different equations of
motion for each method. This also results in specific limitations and
advantages in each case.

Dissipative Particle Dynamics (DPD), the method which is probably closest to 
traditional molecular dynamics, is an off-lattice method which has 
the capability to resolve hydrodynamic space and time scales 
significantly larger than those in traditional MD.
The particles in DPD represent whole atoms or regions of solvent atoms and  
interact via effective forces which conserve momentum locally and 
deliver the correct hydrodynamic behaviour even for relatively low particle
numbers. DPD integration algorithms slightly differ from conventional
MD integrators and take into account the specific symmetry requirements 
in order to fulfil local momentum or global energy conservation.
DPD is particularly well suited for specific physicochemical interactions on 
the particle-level, e.g., for systematic coarse-graining of solvation effects, 
heat conduction and convection in nano-materials, or the simulation of 
biological membranes, macromolecules or multiphase systems
subject to different flow conditions and geometries.

Multi-particle collision dynamics (MPC) implements discrete streaming and
collision steps to update the positions and velocities of an ensemble
of fluid particles. The molecular interactions are coarse-grained by
sorting the particles into collision cells and applying a simplified
collision rule to exchange momentum. The collision cells also function
as averaging volumes to calculate local hydrodynamic fields. This
makes MPC amenable to a kinetic description from which the
hydrodynamic Navier-Stokes equations emerge. In this sense, MPC
bridges between the particle view and the kinetic view of the
fluid. The possibility to tweak the collision rule such that different
thermodynamic ensembles can be reproduced make MPC well suited to
study transport phenomena under different ambient conditions. Heat
transport due to temperature gradients and other non-equilibrium
effects are important examples for the use of MPC.

The lattice Boltzmann method (LBM) is a lattice-based scheme that discretises time, 
space and velocity space. The commonly chosen discretisation leads to a simple and 
nearly local algorithm that naturally lends itself to parallel simulations.
Due to its kinetic nature, complex and moving boundary conditions in the LBM are 
relatively straightforward to implement, when compared to conventional Navier-Stokes solvers.
Since LBM operates with averaged particle distributions, it is especially suited for 
non-Brownian problems, e.g., non-colloidal particle suspensions, although fluctuations 
can be re-introduced. LBM is a popular method for multi-phase or multi-component problems, 
e.g., droplet dynamics or water/oil flow in porous media.
Furthermore, it is straightforward to couple additional physics with the LBM algorithm, 
which enables applications such as liquid crystal dynamics or electrophoresis.

\rev{Interestingly, direct comparisons between the three above mentioned methods are scarce
-- at least we are not aware of any. This is perhaps a knowledge gap that the computational 
soft matter community may want to address in the future. Nevertheless, it is possible 
to draw a comparison between some more general aspects of each method.\\
A generic property of DPD is that both solute and solvent are modelled through  
coarse-grained particles that resemble each other closely. This characteristic feature 
makes the inclusion of specific solute-solvent interactions simple. These 
can also be modelled via effective coarse-grained force-fields that describe the averaged 
atomistic or molecular interactions on a larger length and time scale. 
Both DPD and MPC are fully thermalised methods and are natively connected to typical 
thermostatic or thermodynamic algorithms like Nos\'e-Hoover thermostats or Langevin dynamics. 
Whilst this is usually seen as an advantage, it allows only for fluctuating solutions of hydrodynamic
problems. While fluctuations can be incorporated as well into LBM, arguably in a slightly more 
complicated manner, LBM permits primarily quasi noise-free, ballistic solutions. In form of the 
Chapman-Enskog expansion, LBM offers a tool which allows all specific modifications 
of the microscopic dynamics to be directly related to a macroscopic target equations, providing 
quasi- or near-analytical insight into physical problems.}

Mesoscopic modelling continues to be an area of active research, and
multiphase fluids and non-equilibrium soft matter offer a range of
interesting research problems. Moreover, we anticipate that mesoscale
methods will play a central role in enabling extreme-scale
applications on future HPC systems. We hope that this tutorial review
will help the reader to choose the appropriate methods to address the
mesoscale physics that is relevant in their research problem.

\section*{Acknowledgments}

The authors are indebted to Peter V. Coveney for his support and many valuable discussions. OH acknowledges support from the EPSRC Early Career Fellowship Scheme (EP/N019180/2). TK thanks the University of Edinburgh for the award of a Chancellor's Fellowship.

\footnotesize{
\bibliography{smtutrev} 
\bibliographystyle{rsc} 
}

\end{document}